\documentclass[graybox]{svmult}

\usepackage{mathptmx}       
\usepackage{helvet}         
\usepackage{courier}        
\usepackage{makeidx}         
\usepackage{graphicx}        
\usepackage{multicol}        
\usepackage[bottom]{footmisc}
\usepackage{amsmath} 
\usepackage{mathtools} 
\usepackage{xcolor}
\usepackage{comment}
\usepackage{tikz-cd}
\usepackage{amssymb}
\usepackage[T1]{fontenc}
\usepackage[utf8]{inputenc}
\usepackage[american]{babel}
\usepackage[font=small,labelfont=bf]{caption}
\usepackage{tikz}\usetikzlibrary{shapes,arrows,calc}

\makeindex             

\begin{document}
\title*{A survey of statistical learning techniques as applied to inexpensive pediatric Obstructive Sleep Apnea data}
\titlerunning{A survey of techniques for OSA data}
\author{Emily T. Winn, Marilyn Vazquez, Prachi Loliencar, Kaisa Taipale, Xu Wang, and Giseon Heo}
\authorrunning{Winn et al.} 

\institute{Emily T. Winn \at Division of Applied Mathematics, Brown University, Providence, RI, USA, 02912. \email{emily\_winn@brown.edu}
\and Marilyn Vazquez \at Mathematical Bioscience Institute, Ohio State University, Columbus, OH,  USA, 43210. \email{vazquezlandrove.1@osu.edu}
\and Prachi Loliencar \at Department of Mathematical and Statistical Sciences, University of Alberta, Edmonton, AB, Canada, T6G 2C1. \email{lolienca@}ualberta.ca
\and Kaisa Taipale \at C.~H.~Robinson, 14701 Charlson Rd, Eden Prairie, MN, USA. \email{kaisa.taipale@chrobinson.com}
\and Xu Wang \at Department of Mathematics, Wilfrid Laurier University, Waterloo, ON, Canada, N2L C35. \email{xwang@wlu.ca}
\and
Giseon Heo (corresponding author) 
\at  School of Dentistry; Department of Mathematical and Statistical Sciences, University of Alberta, Edmonton, AB, Canada,  T6G 1C9. \email{gheo@ulberta.ca}}
%
%
\maketitle

\abstract{Pediatric obstructive sleep apnea affects an estimated 1-5\% of elementary-school aged children and can lead to additional health problems. Swift diagnosis and treatment are critical to a child's growth and development, but the variability of symptoms and the complexity of the available data make this a challenge. We take a first step in streamlining the process by focusing on inexpensive data from questionnaires and craniofacial measurements. For exploratory data analysis, we apply correlation networks, the Mapper algorithm from topological data analysis, and singular value decomposition. We then apply a variety of supervised and unsupervised learning techniques from statistics, machine learning, and topology, ranging from support vector machines to Bayesian classifiers, density-based and graph-based clustering, in an effort to characterize and classify pediatric obstructive sleep apnea. Finally, we compare the results of each of these methods and discuss the implications for a multi-data-sourced algorithm moving forward.}


\section{Introduction}
Obstructive sleep apnea (OSA), a form of sleep-disordered breathing characterized by recurrent
episodes of partial or complete airway obstruction during sleep, is a serious health problem, affecting an estimated 1-5\% of elementary school-aged children \cite{OSA1, OSA2}. Even mild forms of untreated pediatric OSA may cause high blood pressure, behavioral challenges, or impeded growth. Compared to adults, the symptoms of childhood-onset OSA are more varied and change continuously with development, making diagnosis a difficult challenge. The complexity of the data from surveys, biomedical measurements, 3D facial photos, and time-series data calls for state of the art techniques from mathematics and data science.

Clinical data, including that considered in confirming or ruling out a diagnosis of pediatric OSA, consist of high-dimensional multi-mode data with mixtures of variables of disparate types (e.g., ordinal and nominal categorical data of different scales, interval data, time-to-event and longitudinal outcomes) also called mixed or non-commensurate data. These data obtained from multiple sources are commonplace in modern statistical applications in medicine and health, with thousands, even millions, of features recorded simultaneously from each object or individual.  

In this paper, we analyze symptom data provided by patients and clinicians, while in related work, we analyze polysomnography signals (physiological time series data). These two papers are case studies in a larger project of building an algorithm to aid clinicians in their treatment decisions for pediatric OSA patients. To overcome the difficulties in analyzing high-dimensional multi-mode data from multiple sources, we propose adopting a hybrid approach that interactively combines statistics, computational topology and deep learning to take advantage of their strengths and mitigate their weaknesses.  Statistics provides a suite of tools for model specification and identification, including model estimation and inference, which oftentimes entail a high computational cost.  Computational topology via persistent homology aims to detect `true' signals in high-dimensional data with respect to a varying model parameter (\cite{Edelsbrunner2002Topological}, \cite{Zomorodian2005ComputingHomology}), which contrasts with the conventional statistical approach of estimating one or more model parameters that, at best, yield signals; however, the lack of a coherent approach to statistical inference in topological data analysis is a serious drawback.  Deep learning, on the one hand, has been successfully used in speech recognition \cite{8683453} and owes part of its practical appeal to its computational efficiency; on the other hand, it can be difficult to intuitively justify why deep neural networks work. Integrative ensemble methods would thus ideally blend statistical theory, topology and deep learning in a seamless fashion, all three working in concert, as in a musical ensemble, fully exploiting the amount of available information across multiple sources.

Here we illustrate the first step toward to building a hybrid approach by comparing several methods in statistics, computational topology, and machine learning. In Section \ref{sec:data}, we describe data collected from pediatric OSA study Pro00057638, at the University of Alberta. Survey and craniofacial data are easier and cheaper to collect than polysomnography (PSG) data, so for maximum clinical impact we focus this article on analyses of those data sets. For the analyses of PSG time series data, we refer the reader to \cite{WiSDM_PSG_2019}.
 
The rest of the paper is organized as follows: in Section \ref{sec:data}, we introduce two sets data -- survey questionnaires and craniofacial scores.  In Section \ref{sec:explore}, we outline initial findings of our data exploration which provides a basis for the methods we applied to our data, described in Section \ref{sec:methods}. In Section \ref{sec:results}, we compare methods in statistics and machine learning for classifying OSA patients using survey data and craniofacial data and discuss results. Lastly, we complete our article with conclusion and future research steps in Section \ref{sec:conclusion}.

\section{Pediatric Obstructive Sleep Apnea and Data}\label{sec:data}
 As in adults, OSA in children is associated with cardiovascular dysfunction, neurocognitive dysfunction, behavioral issues, and metabolic consequences. OSA is also believed to negatively influence school performance and learning potential in children. The gold standard for diagnosis of pediatric OSA is PSG~\cite{PSG}. However, in many countries, access to PSG is severely limited and many children do not have an appropriate diagnosis before treatment.  Consequently, children with OSA may not be treated or some children without OSA may undergo unnecessary surgery. A simple and accessible way to identify children with OSA is needed. Finding insights within inexpensive data is a crucial first step.
 
In our study, patients at risk of OSA underwent PSG, filled out questionnaires, and had 3D photos taken, which were assessed by orthodontists for craniofacial index. Normative patients (considered not at risk of OSA) did not undergo PSG, but filled out questionnaires and were assessed by orthodontists in the same way as patients at risk of OSA. Our analysis of PSG data can be found in \cite{WiSDM_PSG_2019}. 

Broadly speaking, data types can be distinguished as structured and unstructured.
Examples of the former include metabolite concentrations, medical records, and survey questionnaires, while more complex high-dimensional data such as digital images (e.g., photos, CT scans, MRI), text, time series, audio, and DNA sequences are examples of the latter. Methodologies for analysis differ based on whether data are structured or unstructured.
In following sections we demonstrate methods in statistics, computational topology and deep learning to classify or cluster OSA patients using survey questionnaires and craniofacial data (both structured). Our current research aims to build a foundation which will be useful for combining analytic methods for both structured and unstructured data in predicting severity of OSA.

\subsection{Survey Data} \label{sec:survey}

Once a clinician suspects that OSA may cause troublesome symptoms in a child, OSA-specific surveys can be administered to the affected parents and child. The questionnaires analyzed here encompass the Child's Sleep Habits questionnaire, the OSA-18 Quality of Life survey, a Health Screening Questionnaire, the Pediatric Sleep Questionnaire, and the PedsQL Pediatric Quality of Life Inventories for child and parent. 
In the case of pediatric surveys, many children are not old enough to read or respond to such a survey, leaving parents to report observations about symptoms as best as they can. Missing data may result from survey-takers not knowing an answer to a question or feeling uncomfortable answering a question truthfully. However, even with their shortcomings, surveys are far easier and less costly to obtain and analyze than a PSG exam. Usually, PSG exams are not covered by insurance, resulting in out-of-pocket costs; they require a separate appointment and overnight evaluation; and the results can take anywhere between six months and two years to come back. Conversely, surveys can be completed at the clinic during a visit, are of no additional costs to patients or their families, and can be evaluated immediately. 
In addition to the ``subjective'' patient -- and parent -- provided data about symptoms and quality of life, we include dentist-gathered data about craniofacial characteristics (CF data) of children where noted below. 

\subsection{Craniofacial Data} \label{sec:dataCF}
 Of particular interest to clinicians is craniofacial data \cite{CFOSA2017}, which is a series of measurements taken to capture the shape of the face and mouth. There are two reasons for a potential preference for craniofacial data instead of survey or PSG data. First, craniofacial data is inexpensive and takes minutes to measure, and therefore takes up far fewer resources than those needed for a PSG. The accessibility of this data is demonstrated in our data set, which is complete for all craniofacial measurements, eliminating the need for imputation of data. Second, craniofacial data consists of quantitative measurements, which may reduce bias that may arise in responses to qualitative survey questions. 

\begin{figure}
    \centering
    \includegraphics[scale = 0.5]{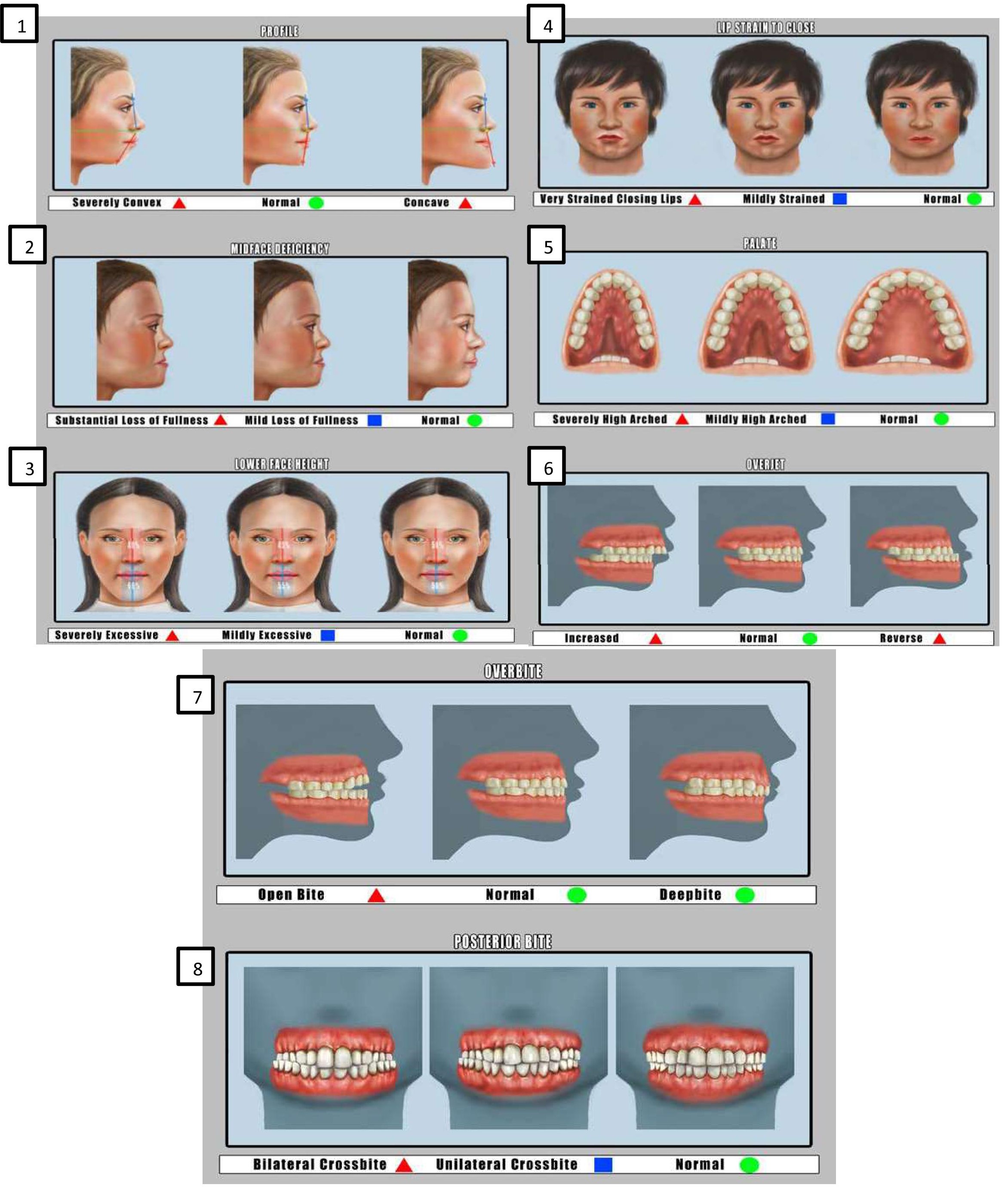}
   \caption{Table illustrating eight of the measurements taken for craniofacial data. The image is taken from \cite{CFindex2014}. A green circle receives a numerical score of 0, a blue square receives a numerical score of 1, and a red triangle receives a numerical score of 2. The ninth score, the Craniofacial Index, is the sum of these eight measurements.}
    \label{fig:CF_illustrate}
\end{figure}

There are nine craniofacial measurements we consider in our analysis. \textbf{Figure \ref{fig:CF_illustrate}} depicts the first eight features, while the ninth, the Craniofacial Index, is a sum of these first eight measurements. The measurements are defined as follows \cite{CFindex2014}:

\begin{enumerate}
    \item \textbf{Profile} is a measurement for the angle of the shape made by the line from the brow to the base of the nose and a line from the base of the nose to the chin when viewing the patient from the side. 
    \item \textbf{Midface Deficiency} quantifies the projection of the malar area below the eyes (the bones which form the eye socket and cheekbones) relative to the rest of the face.
    \item \textbf{Lower Face Height} is the proportion of the length from the brow to the base of the nose to the length from the base of the nose to the bottom of the chin. 
    \item \textbf{Lip Strain} scores the amount of effort a patient uses to close their lips, measured by observing the muscle contractions from the front view.
    \item \textbf{Palate} scores the depth of the palate (the roof of the mouth) and the arch of the palate.
    \item \textbf{Overjet} is the horizontal distance between the upper incisors and the lower incisors when the patient bites down. Any measurement above 5 mm is considered severe.
    \item \textbf{Overbite} is the length of vertical overlap between the upper and lower incisors.
    \item \textbf{Posterior Bite} is the transverse relationship between the molars and premolars, assessed by observing the relationship between upper posterior teeth and lower posterior teeth on both sides of the mouth. 
    \item \textbf{Craniofacial Index}
    is the summation of the previous eight scores and gives a summary statistic of the craniofacial data.
\end{enumerate}

\begin{figure}
       \begin{tabular}{cc}
       \includegraphics[scale = 0.5]{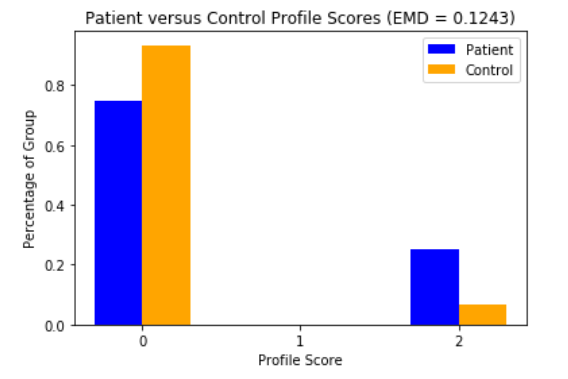} 
       & \includegraphics[scale = 0.5]{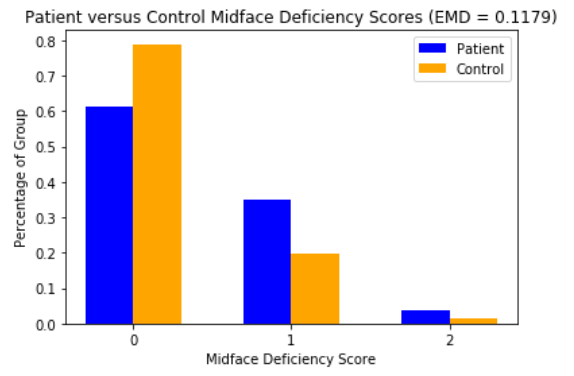} \\
    \includegraphics[scale = 0.5]{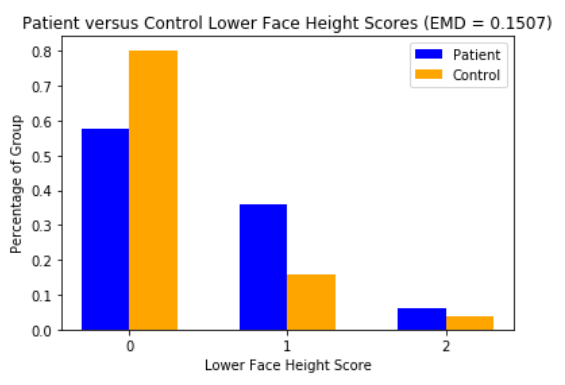} 
       & \includegraphics[scale = 0.5]{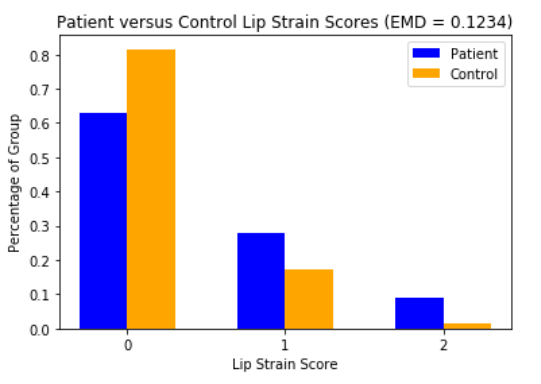} \\
       \includegraphics[scale = 0.5]{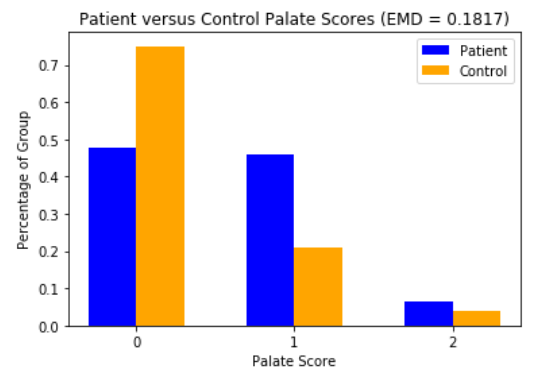} 
       & \includegraphics[scale = 0.5]{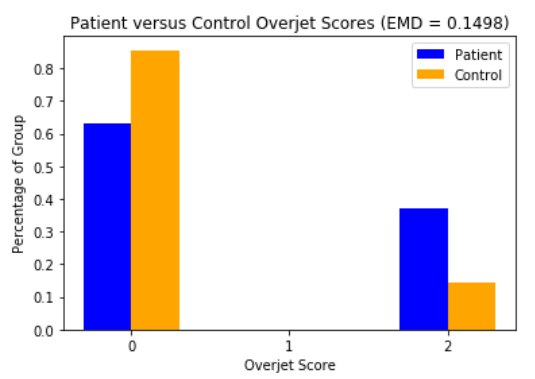} \\
       \includegraphics[scale = 0.5]{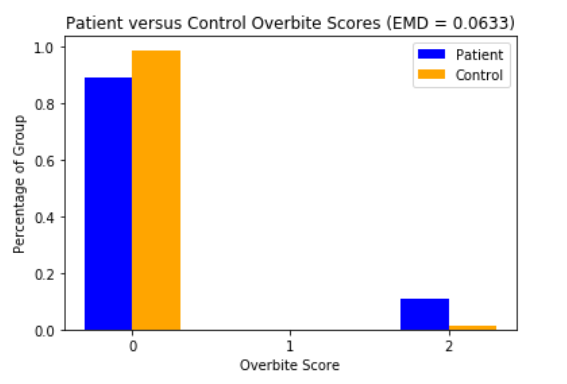} 
       & \includegraphics[scale = 0.5]{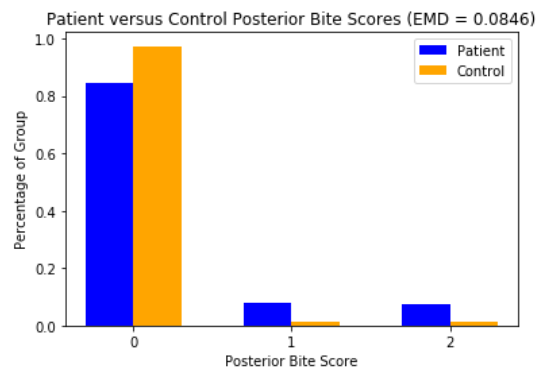} \\
   \end{tabular}                
    \caption{Distributions of different craniofacial measures for the control and the patient groups. Graphs are scaled for the  overall frequency in the data sets. We include the Earth Mover's Distance (EMD) in the title of each feature's graph to quantify the difference between distributions to compare among craniofacial features. As shown, the distributions which are most different are Lower Face Height and Palate Score. The full frequency numbers and Earth mover's distances can be found in Table \ref{tab:CF_summary} in the appendix.}
    \label{fig:CF_graphs}
\end{figure}

\begin{figure}
    \centering
    \begin{tabular}{c c}
        \includegraphics[scale = 0.45]{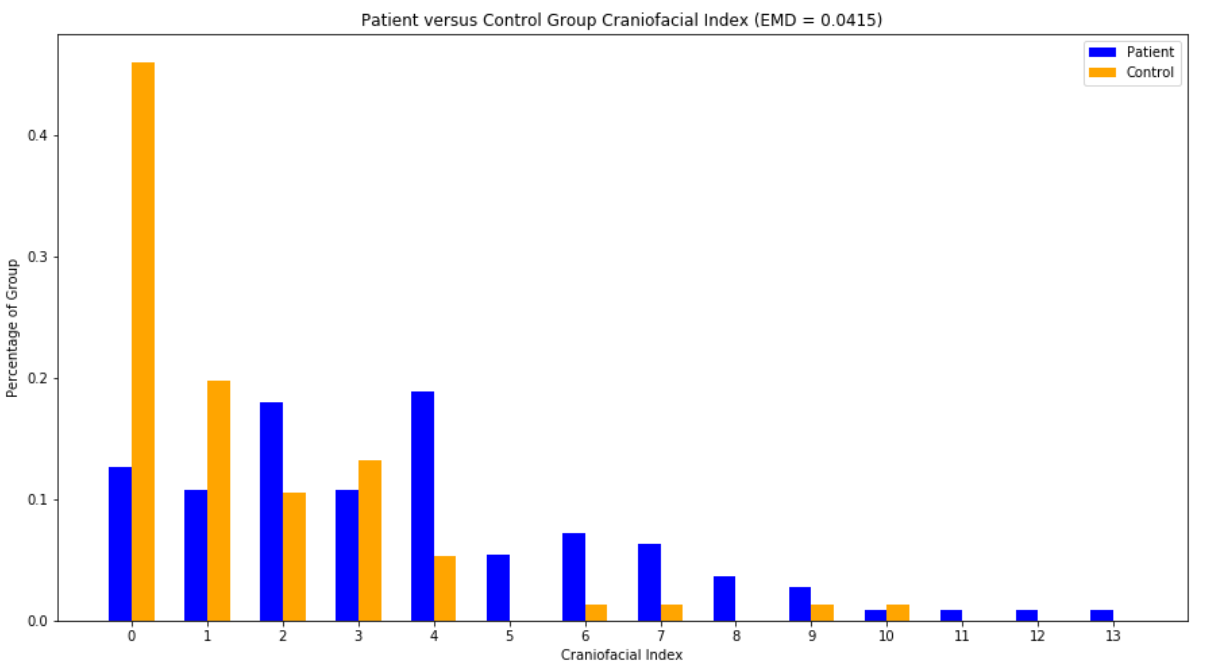}
    \end{tabular}
    \caption{Distribution of Craniofacial Index for the control (right,  orange) and patient groups (left, blue). The graph was rescaled for frequency with respect to the size of their data sets. The Earth Mover's Distance between these two distributions is 0.0415.}
    \label{fig:DTS}
\end{figure}

Measurements 1-8 are scored on a scale from 0 to 2, where 0 indicates a normal measurement and 2 indicates a severely abnormal measurement. As a result, the Craniofacial Index can range from 0 to 16. To analyze the craniofacial data, we first looked at the overall distributions of the complete data set (187 subjects, with 76 controls and 111 patients), visualized as histograms (see \textbf{Figures \ref{fig:CF_graphs} and \ref{fig:DTS}}) to spot any glaring differences between control and patient groups in distributions and calculated the Earth Mover's Distance between distributions for each craniofacial characteristic to quantify those differences (see Table \ref{tab:CF_summary}). The craniofacial characteristics with the most differences were identified. We then conducted our analysis as described in Sections \ref{sec:explore} and \ref{sec:results}.


\subsection{Cleaning Data} 

As with many data sets, ours came with missing values. Before starting our analysis, we cleaned our data set by leaving out samples with too much missing data, imputing missing values where appropriate, and dropping certain variables which also had too much missing information. Our data consists of survey responses and craniofacial data from 200 subjects from two different clinics, with 172 observed variables. After removing subjects who did not have an OSA classification listed, we had 187 subjects remaining. At the early stage of recruiting normative children, some subjects did not fill out questionnaires due to a miscommunication between the principal investigator and research collaborators. Other subjects were too young to be able to fill out any child surveys and thus had too much missing data. As a result, we removed any subjects with more than 50\% of the variables missing. This brought our total subjects down to 173 (67 controls, 106 patients). We excluded text responses from our analysis, such as lists of medications or descriptions of pain. Yes-no questions were encoded in binary. Bed time and waking time were not considered in this analysis, and we encoded variable ``total sleep time'' in one numeric column. Finally, we removed gender, height, weight, and body mass index (BMI) due to the number of missing values. After these steps, we were left with 157 input variables. 

Many of the questions in the surveys were on a Likert scale. For example, patients were asked to rank on a scale of 1-7 how much they agree with certain statements such as \textquotedblleft I have a hard time getting out of bed'', or \textquotedblleft I feel sleepy during the day''. We standardized encoding so that higher numbers indicate the presence OSA symptoms and low numbers the absence of OSA symptoms. For the K-Mapper algorithm and singular value decomposition (see \textbf{Sections \ref{sec:mapper}} and \textbf{\ref{sec: svd}} respectively), we scaled all data to be in the range of $[0,1],$ using $\frac{x-\mbox{min}(x)}{\mbox{max}(x)-\mbox{min}(x)}$. We did not scale the data for other techniques, as they are all scale invariant 

Imputation was carried out separately for methods that used the numerical encoding for categorical variables versus those that used the categorical variables directly. For the former, we imputed missing values using the \textit{MissForest} command from the \textbf{MissingPy} package for Python \cite{missForest}; these numbers were rounded to the nearest integer to be consistent with the raw data.  For the latter, more specifically for Bayesian methods, we used the MICE package with Predictive Mean Matching in R for its capability of imputing categorical data without encoding. Note that any discretization of continuous variables for these methods was carried out after imputation.

We split the data such that 70\% was contained in the training set and 30\% was in the test set. To demonstrate stability, we applied each learning algorithm to 10 different training/test splits with the same 70/30 ratio, where each training and test split had the same control to patient ratio as the whole dataset. We used the same training and test data sets for each supervised learning algorithm unless otherwise stated. For each split, the optimal parameters were found for each training set before evaluating the algorithm on the test data. Specific validation techniques for each algorithm are discussed in \textbf{Section \ref{sec:methods}}. The algorithms were run on three subsets of data: survey questions only, craniofacial measurements only, and  both survey and craniofacial data combined.

\section{Data Exploration} \label{sec:explore}

In this section we present super-level sets of correlation networks \cite{CorrNet2020}  and visualizations from the Mapper algorithm from topological data analysis \cite{KeplerMapper2019}, which we compare to visualizations from singular value decomposition \cite{trefethen1997numerical}. These explorations of the data illustrate the heterogeneity of symptoms experienced by children and shed light on the limitations of the classification algorithms explored later in Section \ref{sec:methods}. 

As an initial exploration of the data, we plotted super-level sets of correlation networks for questionnaire responses and craniofacial scores of the patient and control groups. 
Networks allow visualization of the relationships between input variables and comparison of those relationships across datasets. Correlation network analyses have been successfully applied in computational biology \cite{corrNetwork16}, neuroscience \cite{BrainCorrNetwork}, and finance \cite{FinanceNet2020}. 
For this particular method, we exclude the children's Pediatric Quality of Life survey, as those values likely correlate strongly to the answers of the parents' Quality of Life survey.

\subsection{Correlation Networks}
\label{sec:corr_networks}

\begin{figure}[h]
    \centering
        \begin{tabular}{c}
            \includegraphics[scale =0.4]{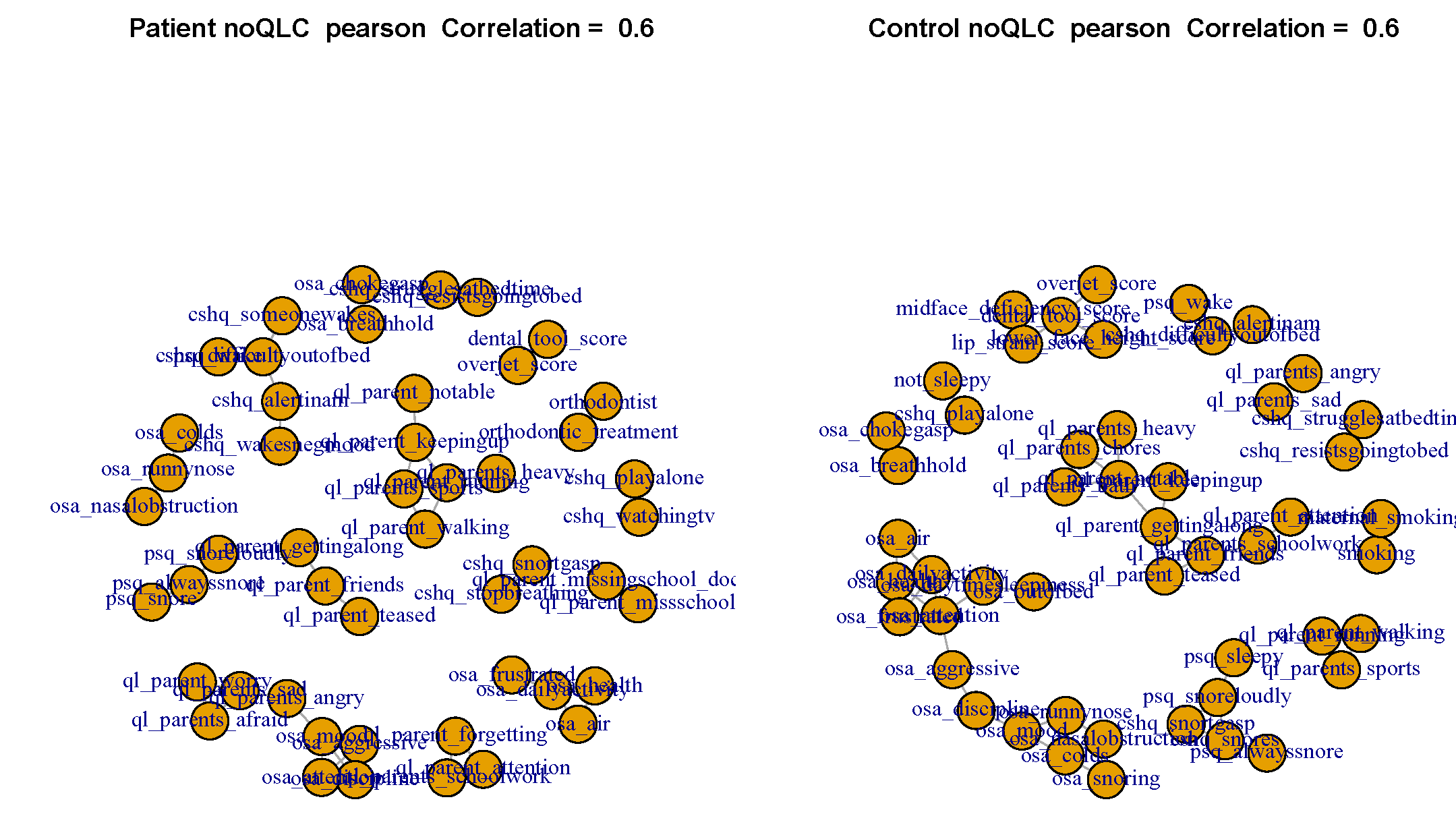} \\
            \scriptsize{(a)} \\
            \includegraphics[scale = 0.4]{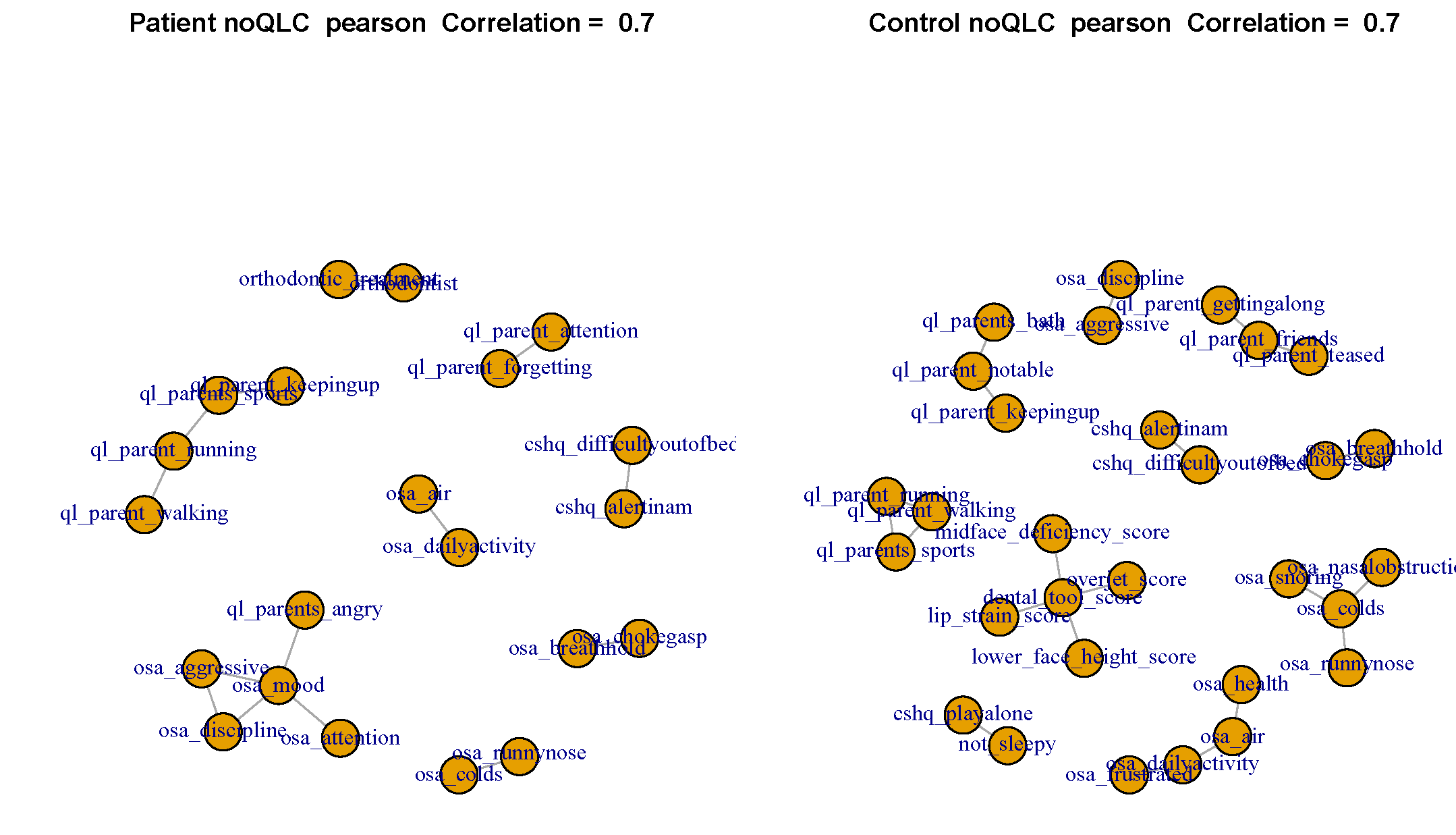} \\
            \scriptsize{(b)}
        \end{tabular}
    	\caption{Correlation networks derived from the raw data. 
    	If two variables had a correlation higher than 0.6  (in (a) top) or 0.7 (in (b) bottom), then an edge was drawn between variables. Isolated points were not plotted. The Quality of Life (QL) Survey answered by children was not included so as to have clearer graphs, as children's answers were likely to be highly correlated with their parent's or guardian's answers to the same survey.}
    \label{fig:CorrGraphs}
\end{figure}
We plot one network representing the patient data and one network representing the control data. Each node or vertex in a network is a symptom/survey response item. For each pair of survey questions $i$ and $j$, we calculate the Pearson correlation between questionnaire responses across all respondents. If the correlation between survey responses $i$ and $j$ is greater than some threshold $h$, then we place an edge between $i$ and $j$. If we consider $h \in [-1,1]$, the networks obtained by varying $h$ form a filtration of the simplicial complex given by the complete graph on all nodes. The topology of these super-level sets gives information about what symptoms are more co-incident in both pediatric OSA and control patients. In \textbf{Figure \ref{fig:CorrGraphs} (a)}, the threshold is $h = 0.6$, while in \textbf{Figure \ref{fig:CorrGraphs} (b)}, the threshold is $h = 0.7$. These values were chosen to illustrate the steps of the filtration. No two variables had a correlation above 0.8, which we attribute to the size of the data set and the variability of symptom expression in OSA patients. To make the plots easier to examine, we only plot nodes with a degree of one or more, so any variables not shown can be assumed not to meet the correlation threshold with any other variable. 


In the network graphs with threshold $h=0.6$, there are some symptom correlations which seem obvious. For example, a patient having seen an orthodontist correlates highly with a patient having received an orthodontic treatment, as one is a prerequisite for the other. However, in \textbf{Figure \ref{fig:CorrGraphs} (a)}, we see that although the patient and control graphs seem to have the same number of nodes (both have 49 nodes in the figure, the rest are isolated), the control group has fewer connected components (11, vs 15 in the patient group, not including isolated nodes). The control group network also has more edges and cliques (53 edges, 12 cliques) compared to the patient group network (38 edges, 6 cliques).  We observe that in this first set of graphs, the control group has higher connectivity than the patient group. 

 This connectivity difference is still present though not as blatant in the second set of graphs in \textbf{Figure \ref{fig:CorrGraphs} (b)}, where the connectivity once again appears stronger in the control group than in the patient group. The control group's network has 29 nodes, 21 edges, and 10 connected components (not including isolated nodes), while the patient group's network has 21 nodes, 14 edges, and 8 connected components. The presence of more edges and more nodes within connected components of the control group's network indicates that some subsets of variables in the control group may have high correlations
 than the same subsets of variables in the patient group. 
 We compare this information with the visualization of patient classification in \textbf{Section \ref{sec: svd}} on singular value decomposition.  Knowing this distribution information is important because it is not only important for clinicians to predict when a child has OSA but also to spot when they do not.  In particular, we notice that five of the craniofacial variables are in a connected component in the control group, whereas all craniofacial variables are isolated nodes in the patient group. This may indicate that craniofacial variables may be useful in finding controls, but not as much valuable for diagnosing OSA. Given the value of craniofacial measures as discussed in Section \ref{sec:dataCF}, we note that the subset of craniofacial measures and their relationship to OSA diagnoses is worth exploring separately from the survey data.

\subsection{Mapper Algorithm} \label{sec:mapper}
The Mapper algorithm from topological data analysis (\cite{Singh2007TopologicalMF}, implemented as \cite{KeplerMapper2019}) is a tool for abstracting high dimensional data so that one can recover the underlying topological structure (the topological nerve of the data). We here refer to the algorithm as the K-mapper algorithm to indicate both the method and the implementation in Python. K-mapper reveals the shape of data via a simplicial complex representation. Given the high dimensionality of the data and the suite of topological and geometric classification methods available, applying the K-Mapper algorithm to our data may indicate whether such geometric and topological methods are worth implementing, or whether we might gain information from those techniques that we would not gain from traditional statistical tools. This methodology has gained some traction elsewhere in exploring medical data (\cite{diabetespaper}, \cite{adhdpaper}).

\begin{figure}
    \centering
        \begin{tabular}{cc}
            \includegraphics[scale =0.25]{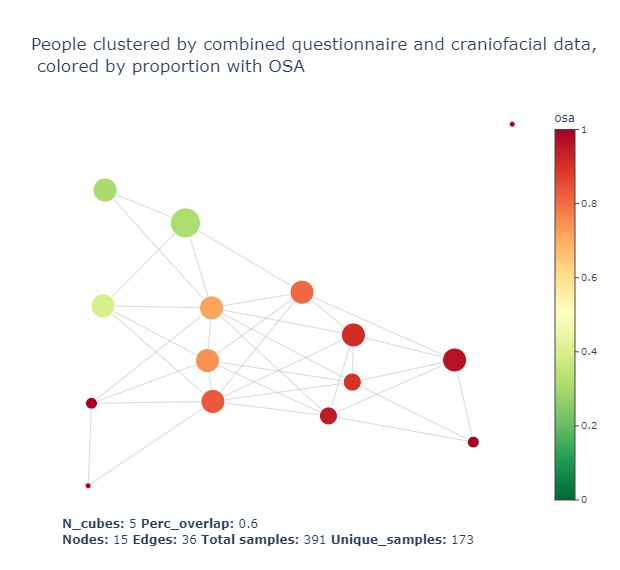} &
            \includegraphics[scale = 0.6]{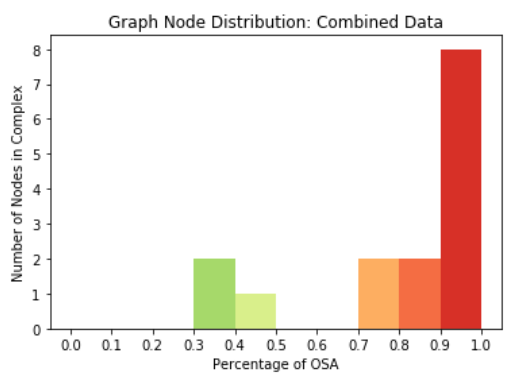} \\
            \scriptsize{(a)} & \scriptsize{(b)} \\
            \includegraphics[scale = 0.25]{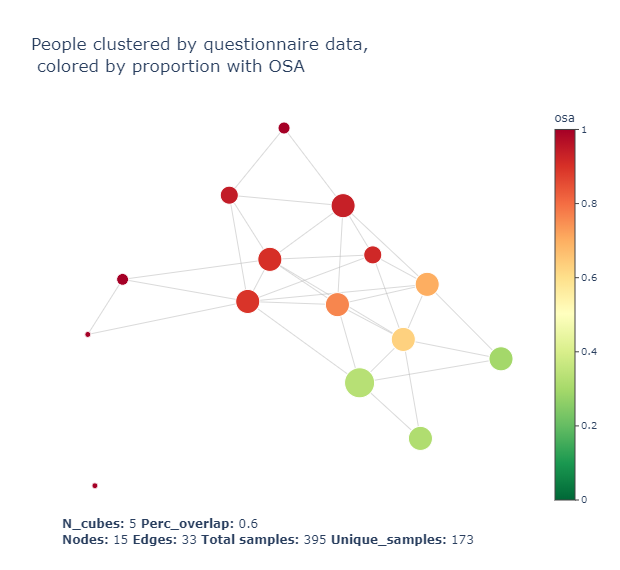} &
            \includegraphics[scale = 0.6]{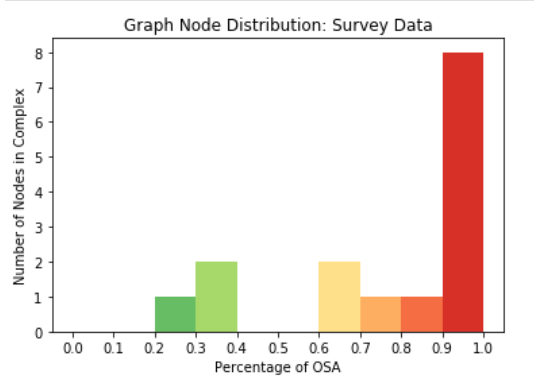} \\
            \scriptsize{(c)} & \scriptsize{(d)} \\
            \includegraphics[scale = 0.25]{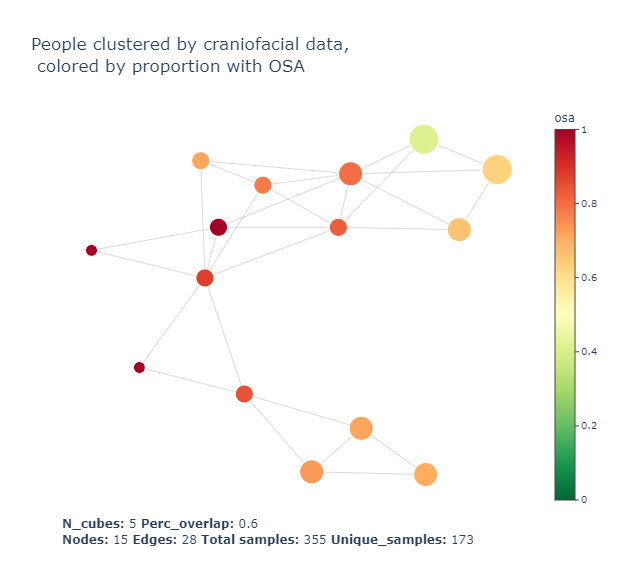} &
            \includegraphics[scale = 0.6]{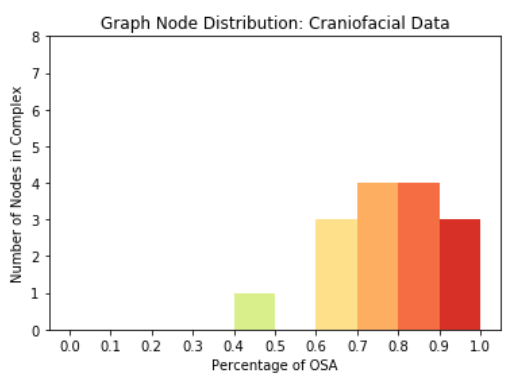} \\
            \scriptsize{(e)} & \scriptsize{(f)}
        \end{tabular}
  
    \caption{Simplicial complexes derived using the K-Mapper algorithm and their respective graph node distributions by color code. Top row: (a) the simplicial complex for combined questionnaire and craniofacial data and (b) the graph node distribution of the simplex, which shows that 8 nodes have over 90\% OSA patients, and the smallest percentage of OSA patients are two nodes in the 30-40\% bin. Second row: (c) the simplicial complex for questionnaire data and (d) the graph node distribution of the simplex. Note the simplex and the distribution are very similar to that for the combined data, but the survey data was able to get one node to be under 30\% OSA. Third Row: (e) the simplicial complex for craniofacial data and (f) the corresponding node distribution. The craniofacial data was not able to parse out the controls as well as the questionairre data, as seen with having only one node below 50\% OSA. However, the craniofacial data itself found 11 OSA patients and put them in clusters of 100\% OSA that the survey and combined data were not able to in their own clustering.}
    \label{fig:KMapper_cheap}
\end{figure}

We briefly describe the Mapper algorithm. First, a filter function is used to define a covering of the data point cloud $D\subset \mathbb{R}^n$, where $n$ is the number of variables under consideration. We used the sum filter function function $f:\mathbb{R}^n \rightarrow \mathbb{R}$, $\vec{x} \mapsto \sum_{i=1}^n x_i$, i.e. $f$ is used to pull back an open cover $\{ U_i \}_{i=1}^5$ of the image $f(D) \subset \mathbb{R}$ given by five equally-sized intervals $U_i \subset \mathbb{R}$, $i=1,...,5$.  After this step, an unsupervised clustering method selected by the analyst is used to cluster data points within the open sets $f^{-1}(U_i)$, and these clusters become nodes of the graph. We used $k$-means clustering with $k=3$. Since each data point can appear in multiple clusters/nodes of the Mapper network, the last step draws an edge between two nodes if there is more than 60\% overlap (this overlap parameter is again set by the analyst).


Topologically, there is not much of a difference between the structure of the simplicial complexes given by combined data (\textbf{Figure \ref{fig:KMapper_cheap} (a) and (b)}), the survey data (\textbf{Figure \ref{fig:KMapper_cheap} (c) and (d))}, and the craniofacial data (\textbf{Figure \ref{fig:KMapper_cheap} (e) and (f))}. While both the combined data simplicial complex and the survey data simplicial complex have two components, the second component is one node made up of exactly one sample. The combined data has more edges (36) than the survey data (33), which has more edges than the craniofacial data (28), but this may very well be a result of the differing number of variables being considered by the algorithm. (Recall there are 157 input variables, 8 of which are craniofacial variables and 149 of which are survey variables.)

However, it appears that the craniofacial data does not have a single cluster with less than 40\% OSA patients, as shown in the coloring of the graphs and in the node distributions, whereas both the survey data and the combined data are able to recover nodes where there are around 30\% OSA patients (the node with the smallest percentage of OSA patients in the combined data has 30.8\% OSA patients, while the node with the smallest percentage of OSA patients in the survey data has 29.6\% OSA patients). This indicates that the craniofacial data by itself is unable to clearly distinguish controls from the collective set of patients and controls, while the presence of survey data grants this distinction among nodes.

That said, the craniofacial data by itself was able to discover slightly more pure nodes which were 100\% OSA, and the samples which fell in these nodes had very little overlap with the 100\% OSA nodes from the survey data, and those of the combined data. In the survey data, there were 4 nodes which were 100\% OSA and together were made up of 8 samples. In the combined data, there were 4 pure OSA nodes, which in total were made up of 9 samples, 8 of which were the same as the those in the pure OSA nodes identified using the survey data. In the craniofacial data, there were 3 nodes made up of total 14 OSA samples, only 3 of which were overlapped with those OSA samples identified by the survey data, and also by the combined data. Thus, the craniofacial data were able to clearly distinguish 11 OSA samples that were mixed in with controls in the other two simplicial complexes. On its own, craniofacial data offers a different angle of consideration for finding those most likely to have an OSA diagnosis. The K-Mapper analysis justifies the need to run algorithms not just on the combined data but on the craniofacial data and the survey data separately, as they seem to give different insights: the craniofacial data is best at finding who most likely has OSA, while the survey data is better at finding who does not have OSA.

\subsection{Singular Value Decomposition}\label{sec: svd}

Singular value decomposition (SVD) is a commonly used matrix factorization method that decomposes a matrix $A$ into a product of three matrices: $A=USV^T$,  where $U$, $V$ are unitary matrices, while $S$ is a rectangular diagonal matrix \cite{trefethen1997numerical}. SVD can be used as a dimension reduction technique through selecting directions of high variance via the left and right singular vectors that make up matrices $U$ and $V$. This method is related to the well-known Principal Component Analysis (PCA). While PCA is a factorization of the covariance or correlation matrix, the SVD offers a full matrix factorization of the original data matrix. Applied to the matrix of respondents and responses, the SVD can suggest the survey questions that are significant in confirming or discarding a diagnosis of OSA. It also suggests ways in which presentation may be divergent among distinct groups of children. Here we examine survey data only, discarding the craniofacial data for the moment. The SVD was applied to the scaled survey data. Here we used the \emph{svd} function included in the base package of \textbf{R}.


\begin{figure} 
\centering
    \begin{tabular}{cc}
            \includegraphics[scale =0.70]{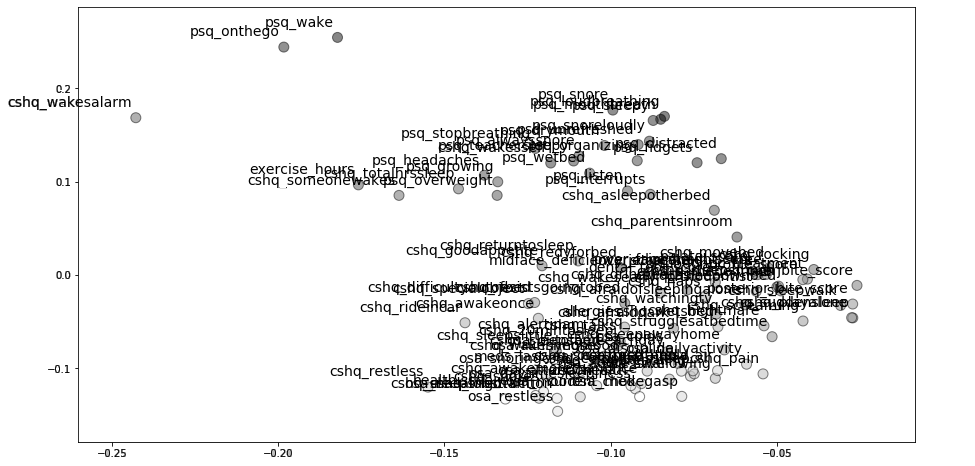} \\
            \includegraphics[scale =0.70]{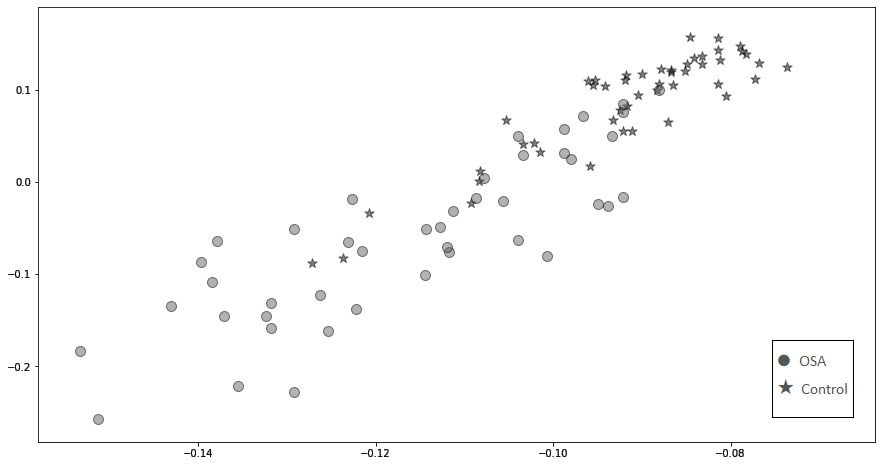} \\
    \end{tabular}
        \caption{Projections to the first two components of the symptom space (top), and projections to the first two components of the respondent space (bottom) as given by singular value decomposition (SVD). The figures underscore the difficult challenge of categorizing patients and prioritizing symptoms. In the top image, the texts are the survey variables. In the bottom graph, the stars denote controls, while the dots are OSA patients.}
        \label{fig:SVD}
\end{figure}

Visualization of the projection to the first two singular components  demonstrates the range of symptoms experienced by participants and illuminates the difficulty that we will see in using supervised learning methods like decision trees and support vector machines to create accurate classifiers. The projections to the first two singular components in symptom space (or variable  space), that is, the first two columns of $U$ are in the top image of \textbf{Figure \ref{fig:SVD}}, with the grayness of the point proportional to the magnitude of average difference between control and symptomatic patients. Meanwhile the bottom graph of \textbf{Figure \ref{fig:SVD}} displays the projections to the respondent space i.e. the first to columns of $V$, with respondents "no OSA" (stars) and "OSA" (dots). As demonstrated in this projection, control patients and symptomatic patients are not cleanly separable, and symptomatic patients are notable for their heterogeneity. This observation foreshadows the poor results we will see from classification algorithms. 

The projection to the first two singular components in symptom space shows that just a few questions exhibit the largest variance in responses: \textit{waking up feeling unrefreshed in the morning}, \textit{waking up with an alarm}, \textit{and having a child who is on-the-go at all times} are high-variance symptoms, with a high difference between control and symptomatic patients. The surprise is that these symptoms do not cleanly differentiate between children with and without OSA. This highlights that pediatric OSA is a complex diagnosis that cannot be easily reduced to appearance of a subset of symptoms.

It is worth comparing the visualizations from singular value decomposition with those coming from kernel PCA, shown in \textbf{Figure \ref{fig:dbc_results}}. 

\section{Statistical Learning Methods} \label{sec:methods}
Because of the high dimensionality of our data and the findings in our data exploration, we apply a variety of machine learning techniques. In this section, we discuss each of the methods used and the justification for applying them to our data sets. The performance results of those classification methods are in Section \ref{sec:results}. We aim to classify recruited patients into two categories: No OSA and risk of OSA. In clinical practice, there are four categories for OSA - no OSA, mild OSA, moderate OSA, and severe OSA - but to establish initial diagnostic results, we start with presence or absence of OSA. Further discussion about classification of severity, using all four categories, is in Section \ref{sec:conclusion}.

\subsection{Non-Bayesian Supervised Learning}\label{sec:survey-supervised}


We applied various supervised methods to see how accurately one can infer whether a subject has OSA or does not have OSA. These  methods include Linear Discriminant Analysis (LDA), Quadratic Discriminant Analysis (QDA), Logistic Regression (LR), Decision Trees (DT), Random Forests (RF), Neural Networks (NNET), and Support Vector Machines (SVM) and K-nearest Neighbour (KNN). These techniques are well-established and we provide only a cursory overview of results so that contrast with other methods can be established. 

Linear Discriminant Analysis (LDA), proposed by Fisher in 1936, is one of the most basic classification methods. The method uses a linear decision boundary, and assumes that the conditional probability density function on each class is Gaussian with a shared covariance matrix for the feature data.  In comparison, Quadratic Discriminant Analysis (QDA) assumes different covariance matrices for different classes, creating quadratic decision boundaries between classes. Although LDA may seem too simple to handle complex high dimensional data, we use it as a benchmark method as it may provide insight on how much improvement more advanced statistical/machine learning methods can achieve.  Logistic regression (LR) is also a very popular method, especially for binary classification and models the probability of classes using a logistic function . We used the R packages \textit{mass} \cite{Rmass} and \textit{glmnet} \cite{Rglmnet} for our calculations.

Decision tree (DT) is a non-parametric method which classifies based on how strongly each individual feature performs in predicting ultimate diagnosis. A major benefit of DT is its interpretability, as it allows the user to see which variables are most significant in the classification. The \textit{DecisionTreeClassifier} from the \textit{Scikit-learn} library for Python was used \cite{scikit-learn}. A grid search to find the optimal cost-complexity pruning parameter was performed, using the model selection library from \textit{Scikit-learn}. The optimized parameters for DT and other machine learning methods can be found in Table \ref{tab:parameterVals}. In addition to DT, we used random forests (RF) to select important variables. Random forests consist of an ensemble of decision trees. Each step growing the component decision trees randomly selects a user specified number of variables from the whole set of variables. The purpose of this randomization is to grow many trees, which are not similar to each other, allowing the algorithm to explore the explanatory variable space as widely as possible, often giving a more robust outcome than decision trees. In this paper, we used the default choice for the number of variables selected at each step, i.e., $\sqrt{P}$, where $P$ is the number of variables. We used the R package \textit{randomForest} for these calculations \cite{RrandomForest}.

Neural Networks (NNET) are the basic component of modern deep learning methods. 
The model creates a graph consisting of multiple "hidden" layers, each of which consists of a specified number of nodes or hidden units. The initial layer consists of the input variables while the final layer models the outcome. Values of nodes in each hidden layer are obtained by taking a linear combination of the features in the previous layer and applying a non-linear activation function (usually the logit function). Using a non-linear activation builds complexity and allows flexible modeling. NNET finds the ideal weights in the linear combinations by iterating between forward and backward propagation to minimize the desired loss function using gradient descent.  While deep learning, entailing a large number of hidden layers and/or units has gained popularity, literature applying it to OSA detection has been limited to adult OSA and does not generally deal with survey or craniofacial data. For example, see \cite{dey} \cite{apneaseverity} for apnea severity classification using ECG data, \cite{deep} for OSA detection using facial images, and \cite{nnetreview} for a literature review. In contrast, due to the size of our survey data, we only applied shallow neural networks with one hidden layer. This left the following parameters to be tuned: (1) the number of hidden units per layer (size = 2, 3, 4, 5, 6, 7), (2) the random seeds for initial values (2000, 2001, 2002, 2003, 2004, 2005) and the  (3) learning rate (i.e., decay = $e^{-2}$, $e^{-1}$, 1). The R package \textit{nnet} was used in our implementation \cite{Rnnet}. 10-fold Cross-validation on each of the training data sets  was used to select the optimal combined set of tuning parameters, and then the corresponding test data set was used to obtain the performance measurements. \\

\begin{table}[]
    \centering
    \begin{tabular}{c}
    \textbf{Example Optimal Parameters for First Train/Test Split for Each Data Set}  \\
\end{tabular}
    \begin{tabular}{p{2cm}p{2.5cm}p{2.5cm}p{2.5cm}}
        \hline\noalign{\smallskip}
        Method & CF Data & Survey Data &  Combined Data\\ \noalign{\smallskip}\svhline\noalign{\smallskip}
        DT & $\alpha = 0.01$ & $\alpha=0.04$ & $\alpha = 0.04$\\ 
        RF & $\sqrt{p} \approx 3$ & $\sqrt{p} \approx 12$ & $\sqrt{p} \approx 13$ \\
        NNET & size = 2 & size = 2 & size = 2\\
        & decay = 1.0$e-2$ & decay = 1.0$e-2$ & decay = 1.0$e-2$ \\
        SVM & kernel - exp. & kernel - linear & kernel - exp. \\
        & C = 1 & C=0.001 & C=1 \\
        & $\gamma = 1$ &  & $\gamma = 0.01$ \\ 
        $k$-NN & $k=9$ & $k=6$ & $k=5$ \\ 
        \hline\noalign{\smallskip}
    \end{tabular}
    \caption{Optimal parameters for supervised learning methods for the first train/test split. For the Decision Trees (DT), $\alpha$ is the cost-complexity pruning parameter. For Random Forests (RF), the $\sqrt{p}$ is the square root of the total number of random variables being considered. For Neural Networks (NNET), ``size" refers to the number of hidden units and ``decay" is the learning rate. The parameters were chosen using the 10-fold Cross Validation of the training data. For Support Vector Machines (SVM), we have three parameters: the type of kernel, the penalty parameter $C$, and the scaling parameter $\gamma$. Finally, for $k$-Nearest Neighbors ($k$-NN), $k$ is the number of data points within the closest proximity of each data point that they are connected to in order to form clusters.}
    \label{tab:parameterVals}
\end{table}

In Support Vector Machines (SVM), a decision boundary is chosen by solving a convex optimization problem. In particular, the method finds a separating hyperplane maximizing the margin between two classes.  The key strength of SVM is allowing the replacement of the Euclidean inner product in the objective function with an arbitrary kernel function, thereby enabling versatile non-linear decision boundaries. Tunable parameters for this algorithm include the penalty parameter $C$ for an $L_2$-regularization term, and a scaling parameter $\gamma$ which determines the proximity of training vectors that will influence the classification of a new point (higher values indicate closer points are used). We optimized these values for linear and exponential kernels (where kernel type was also a variable) using a cross validation grid search. SVM has given promising results in some diagnostic contexts, for instance in diagnosing Lyme disease (\cite{SVMLyme}).


Last, we used $k$-nearest neighbors classification (KNN), which simply put, uses the votes of the $k$ nearest neighbours of a test point to determine its class.
We used uniform weights for each feature as we did not wish to introduce feature importance in a benchmark measure. We used a cross validation grid search to find the optimal $k$ over values $\{2,3,...,9\}$. The \textit{scikit-learn} \textit{KNeighborsClassifier} toolbox for nearest neighbors was used in these calculations \cite{scikit-learn}.

\subsection{Bayesian Classifiers}\label{sec:bayesian}


Bayesian networks are directed acyclic graph models that represent the joint distribution of a set of random variables. In these models, random variables are represented as nodes and dependencies between them are represented by directed edges.
Advantages of Bayesian networks include: their suitability for small and incomplete data sets, their ability to combine known knowledge/priors with data (allowing incorporation of expert knowledge), and avoiding a parametric approach that makes strong assumptions on the data structure (see more details in \cite{sucar} and \cite{uusitalo}). The main disadvantage of these networks, however, is the computational complexity in constructing network structure from data.

Bayesian classifiers are a particular application of Bayesian networks - they estimate the probability of each class given the predictor variables.  We apply two basic classifiers, Gaussian Na\"{i}ve Bayes~(GNB) classifier and Tree augmented Bayesian~(TAN) classifier, and a more recent approach, Semi-Hierarchical Bayesian (SHNB), described in \cite{shnb}.
GNB assumes that all the attributes are independent given the class variable, and further that the conditional distribution of the continuous features is Gaussian for each class. This classifier can be efficient due to the low number of required parameters and the low computational cost for inference and learning. However, in many applications to real life data, conditional independence may not be valid. To help this, TAN constructs a directed tree among the attributes to incorporate dependencies between them, allowing each attribute to depend on other attributes as well as the class variable.

As the number of attributes increases, the number of possible structures goes up, requiring a large dataset to obtain good estimates. To overcome this, the authors in \cite{Zhang2004LatentVD} introduced the Hierarchical Naive Bayesian (HNB) classier, a variation of which is SHNB \cite{shnb}.  SHNB extracts latent ("hidden") variables from highly dependent observed variables and uses these to replace the original, thereby reducing the number of attributes. The analysis is conducted in two steps: 1) creating latent variables using NB, 2) using TAN with the latent variables and the remaining attributes (we remark here that latent variables were created by discretizing continuous variables and applying Naive Bayes, not GNB, as the extension of GNB to TAN or SHNB is not straightforward). To create latent variables, we calculated the conditional mutual information~(CMI), a similarity metric, between each pair of variables given the class.  We then obtained an undirected graph connecting all the attributes based on the CMI between each pair of variables. Lastly, we created latent variables from the variables that form maximal cliques. For example, among 8 craniofacial features, two maximal cliques were chosen with the threshold similarity measure: the first being overjet and overbite, and the second being midface deficiency and lower face height. We created latent variables from each pair, calling them \emph{anterior teeth coupling} and \emph{maxillomandibular facial proportion}, respectively.   The original four attributes consisted of three levels each, displayed in Table \ref{tab:latentCat}.  We combined the levels "deep bite" and "open bite" for the variable overbite and called it \emph{abnormal} (similarly the abnormal category was used for both "increase" and "reverse" levels for overjet). The latent variables thus obtained from each pair consists of 4 categories: Abnormal-Normal, Normal-Normal, Normal-Abnormal, and Abnormal-Abnormal. The SHNB structure was then obtained using the TAN classifier with these latent variables and the four remaining craniofacial attributes that were not a part of any maximal clique. 
\begin{table}
\begin{tabular}{p{2cm}p{1.5cm}p{1.5cm}p{1.5cm}p{1.5cm} p{3cm} p{1.5cm}p{1.5cm}}
\hline\noalign{\smallskip}
OB$\backslash$ OJ & Increased  & Normal & Reverse \\
\noalign{\smallskip}\svhline\noalign{\smallskip}
Deep bite  & 15 & 27 & 0\\
Normal & 15 & 96 & 8\\
Open bite & 3 & 2 & 7\\
\noalign{\smallskip}\hline\noalign{\smallskip}
\end{tabular}
\quad\quad
\begin{tabular}{ccc}
\hline \noalign{\smallskip}
ATC & Normal & Abnormal\\
\noalign{\smallskip}\svhline\noalign{\smallskip}
Normal & 96 &  23\\
Abnormal & 29 & 25\\
\noalign{\smallskip}\hline\noalign{\smallskip}
\end{tabular}
\caption{(Left) Original 9 categories~(cardinalities) of combined OB and OJ. (Right) A latent variable, anterior teeth coupling~(ATC)  with 4 categories.}
\label{tab:latentCat}
\end{table}

We similarly modeled GNB, TAN and SHNB for two additional datasets: all survey questionnaires (total of 149 variables), and both survey and craniofacial index data (total of 157 variables).
The six continuous variables in the survey questionnaires were discretized into two groups and cut off at values: age (8 years old), \# of hours of sleep (9 hours), how long waking lasts at night (12 min), \# of days (last week) participating in physical activity (5 days), quality of life rating called OSA-overall (7 out of 10) and the total number of "yes" responses to OSA-related symptoms (8 out of 22). For both of the aforementioned datasets, the undirected graphs based on CMI between variables showed 3 clusters formed by the variables in each of the following questionnaires: Pediatric Quality of Life Inventory (PedsQ, 23 variables), PedsQL by children (23 variables) and OSA-18 (19 variables).
Each questionnaire recorded responses on an ordinal scale, with a fixed number of response levels shared by all questions in the questionnaire.  We created the latent variable by choosing the most frequent answer for each patient to the questionnaire. For example, the OSA-18 questionnaire had 18 questions, with 7 levels of responses each, where patients responded to frequency of symptoms such as "Choking or gasping sounds while asleep" with answers varying from "none of the time" to "all of the time". If a patient's most frequent response to these 18 questions was "all of the time", the latent variable was set to this value. The PedsQ/PedsQL surveys had no continuous variables so that this technique was sufficient. However, the OSA-18 survey had the additional (discretized) OSA-overall variable - we combined this with the most frequent answer to obtain a final latent variable in a manner similar to the craniofacial variables (taking interactions of levels). Overall we had a total of 3 latent variables, one from each questionniare. The results for these methods are presented in Section \ref{sec:results}. 

 To obtain the GNB model, we used the \emph{GaussianNB} and \emph{CategoricalNB} functions (for continuous and discrete variables respectively) from the Naive Bayes toolbox of the \emph{Scikit-learn} library in Python, combining probabilities using the conditional independence assumption.  For SHNB and TAN, we instead used the R packages \emph{bnlearn} \cite{bnlearn} and \emph{gRain} \cite{gRain} to obtain posterior probabilities (inference) and the model (learning).

\subsection{Unsupervised Learning} 
\label{sec:manifold}

In addition to the supervised learning methods described in the previous sections, we considered exploring data using unsupervised learning methods, especially clustering analysis. Since we had the true class labels for our data (OSA or no OSA), we used these to determine parameters for our methods, and to choose labels for clusters that give the higher accuracy.  In the real world, the clinician/statistician would not have this luxury, and would have to analyze the two clusters produced by the methods to decide which label (OSA or no OSA) applies to each. We however take advantage of the known diagnosis to provide evaluation measures comparing the performance of our clustering methods in the best case scenario (best choice of parameters and label permutation).

To cluster the survey respondents, we applied five different clustering approaches:
\begin{enumerate}
\item Density-based spatial clustering of applications with noise (DBSCAN) \cite{ester1996density}
\item Cut-Cluster-Classify~(CCC) \cite{vazquez2018}
\item  Spectral clustering \cite{von2007tutorial}
\item Continuous k-Nearest Neighbour approach (CkNN) \cite{berry}
\item Thresholding density using $q_k$ (see subsection \ref{thresh})
\end{enumerate}
Among these five clustering approaches, DBSCAN and CCC are density-based clustering methods, while spectral clustering and CkNN are graph-based clustering methods. The tuning of each method's parameters was done by comparing the F1-score (also known as the Dice coefficient), i.e., choosing parameters that gave the highest F1-score measure.


\subsubsection{Density-based spatial clustering of applications with noise (DBSCAN)}
\begin{figure}
    \centering
    \begin{tabular}{cc}
     \includegraphics[width=2.2in]{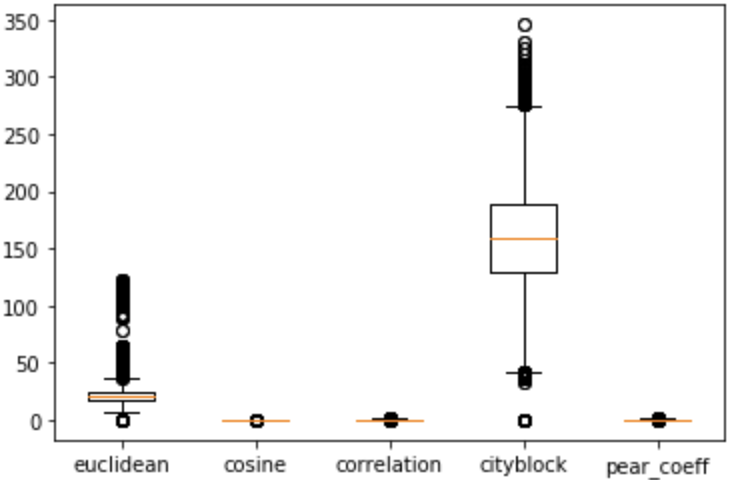} &  
     \includegraphics[width=2.2in]{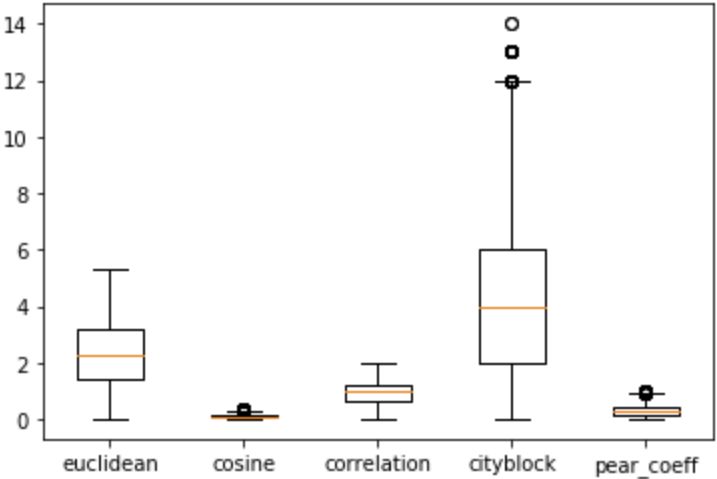}\\
     \scriptsize{(a)}   & \scriptsize{(b)} \\
     \end{tabular}
     \begin{tabular}{c}
     \includegraphics[width=2.2in]{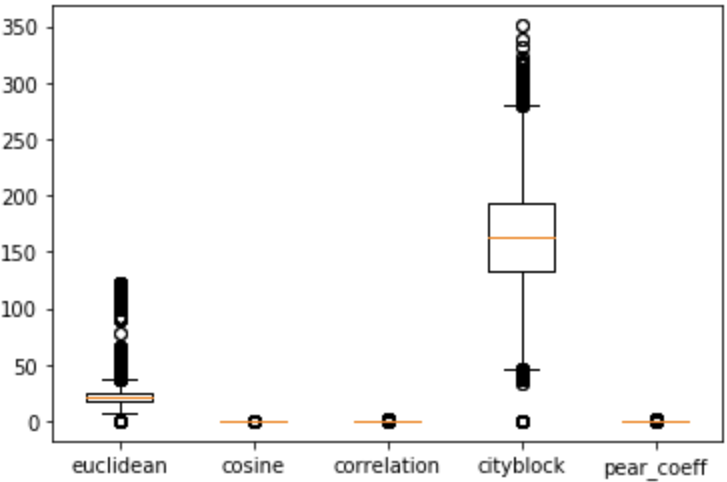} \\
     \scriptsize{(c)} 
    \end{tabular}
    \caption{The distribution of distance values given the different metrics for (a) survey data, (b) craniofacial data, and (c) combined data. }\label{fig:distances}
\end{figure}

DBSCAN is a clustering method that finds neighbourhoods of tightly grouped points i.e., high density regions and assigns them to clusters. The method first finds "core points" that form the center of each cluster, and then connects each core point to its close-by neighbouring points, thereby creating the clusters. Points that fall far from such high density regions are labelled outliers. Since we only wanted two classes, we placed all outliers in the class that gave the largest F1-score. The two main parameters for this method are: \verb|min_samples|, which refers to the minimum number of neighbouring points required to define a core point; and $\epsilon$, the maximum distance allowed between two points in each neighbourhood. We found \verb|min_samples|=2 to give the best F1-score for all methods (since we ran all experiments with the same number of points, this parameter is shared). The value of $\epsilon$ on the other hand, highly depends on the distance metrics, so we studied the distribution of distance values in each distance metric, which can be seen in \textbf{Figure \ref{fig:distances}}. This gave us a good range of values to try, and we again chose the one that gave the optimal F1-score. The optimal values can be seen in Table \ref{tab:epsi}.  

\subsubsection{Cut-Cluster-Classify (CCC)}
\label{cccsub}
The Cut-Cluster-Classify method divides the labeling task into three steps. It starts by picking points that pass a fixed threshold cutoff on the sample density. The points thus obtained are then clustered by connected components. Lastly, the remaining points are classified into one of these clusters. 

The algorithm requires two parameters: the number of points to sample, and the $k$ value used to define the sample density $q_k$:.
\begin{equation}
q_k(x) = \frac{1}{||x-x_k||},
\label{eqn:qk}
\end{equation}
where $x_k$ denotes the $k$th nearest neighbour of point $x$. The parameters that give the  best F1-scores can be found in Table \ref{tab:cknn_and_ccc}(b). 

\subsubsection{Spectral Clustering}
 Spectral clustering exploits the graph structure of data to find an appropriate clustering of its nodes. The classical algorithm for this method starts by constructing a normalized graph Laplacian matrix from similarity measures, then calculates its eigenvectors and uses k-means clustering on these. The only parameter needed is the number of clusters, which for all our experiments is 2.

\subsubsection{Continuous k-Nearest Neighbour approach (CkNN)}
CkNN clustering uses a continuous scale to construct a representative graph of the data. At each scale, it finds a clustering, and then uses consistent homology to choose the best scale. Parameters needed for this method include: the number of points needed to define a neighborhood, and the $k$ value required to define the sample density $q_k$ (\ref{eqn:qk} under subsection \ref{cccsub}). The values with the best F1-scores are given in Table \ref{tab:cknn_and_ccc}(a). 

\subsubsection{Distance Metrics and Thresholding density using $q_k$ }
\label{thresh}
For clustering algorithms, one of the most important issues is which distance metric to use. We compared the F1-scores of clustering methods using five different metrics for discrete points:  Euclidean, Cosine, Correlation, City block (or Manhattan), and 1 minus the absolute value of the Pearson coefficient. The results can be seen in Tables \ref{tab:fscores_survey}-\ref{tab:fscores_all}. Note that the metric giving the best results varied for the different clustering methods. We chose to report the quality metrics for the Manhattan distance in Tables \ref{tab:survey_only},\ref{tab:CF1}, and \ref{tab:combo} since this distance seems to be consistently good for all the methods. \textbf{Figure \ref{fig:dbc_results}} shows some of the labeling results.

Given the hypothesis that the classes differ by sample density, we also estimated the sample density and found a threshold to cluster the data and compare with the other methods. The sample density we used is $q_k$ with a 2-norm (\ref{eqn:qk} under subsection \ref{cccsub}) along with Otsu's method \cite{otsu1979threshold} to do the thresholding. This method analyzes the distribution of values and finds the modes for a threshold value. Results for this method are presented under the shorthand label "Threshold $q_k$" in section \ref{sec:results}. 

\section{Results}
\label{sec:results}

For each method and each data subset, we assess classification success by measuring the accuracy, positive predictive value (PPV, the proportion of subjects classified as having OSA that indeed have OSA), negative predictive value (NPV, the proportion of subjects classified as not having OSA that are controls), sensitivity (the proportion of OSA patients who were classified as having OSA), and specificity (the proportion of controls who were classified as not having OSA). These measures are standard in clinical research literature.  To assess the stability of each method, we used  a ten-split cross validation, or ten different training/test splits of the cleaned data. About 58\% of survey respondents whose responses were used in testing or training sets had diagnosed pediatric OSA; this is our no-information rate. Measures for each method are reported in the form of mean $\pm$ standard deviation in Tables \ref{tab:survey_only}, \ref{tab:CF1}, and \ref{tab:combo}. Unsupervised learning methods had only one measure to report as they do not have training and test splits (see Section \ref{sec:manifold}). These are provided for comparison of methods amongst each other only.   We note that when we applied QDA to the combined survey and craniofacial data, some classes were too small to estimate their respective covariance matrices. Therefore, we did not obtain any results for QDA.

Additionally, for unsupervised methods, we include the Adjusted Rand Index in Tables \ref{tab:ARI_Survey}, \ref{tab:ARI_Cranio}, and \ref{tab:ARI_Combined}. The Rand index is a similarity measure between two clusters that evaluates the frequency of agreements in clustering pairs of data points. The adjusted Rand index is the corrected version of the Rand index that accounts for chance. Here we compare our clustering results to the true class labels.  Assuming our data has $k$ true classes, and our clustering method produces $k$ clusters, we obtain a $k\times k$ contingency matrix $X$, where $X_{ij}$ represents the number of elements of the $i$th cluster that belongs to the $j$th true class. If we represent the sum of the $i$th row by $a_i=\sum_{j=1}^k X_{ij}$, and the sum of the $j$th column by $b_j=\sum_{i=1}^k X_{ij}$, the Adjusted Rand index (ARI) is given by

$$ARI = \frac{\sum_{ij} 
\binom{X_{ij}}{2}-\left[\sum_{i}\binom{a_i}{2}\sum_{j}\binom{b_j}{2}\right]/\binom{n}{2}}{
\frac{1}{2}\left[\sum_{i}\binom{a_i}{2}+\sum_{j}\binom{b_j}{2}\right]-\left[\sum_{i}\binom{a_i}{2}\sum_{j}\binom{b_j}{2}\right]/\binom{n}{2}}$$

\noindent Note that the ARI does not depend on the number of clusters and is invariant under the permutations of the clusters. The score is bounded between -1 and 1, where higher scores indicate better agreement. A score of 1 indicates a perfect clustering,  scores close to 0 indicate a random clustering (no informtion), and negative numbers indicate a performance worse than random clustering.


\subsection{Results for Survey Data}

Of all the classification techniques applied to the survey data, GNB, Random forests (RF), $k$-nearest neighbors performed the most consistently. Unsurprisingly, LR had the lowest accuracy at 0.54$\pm$ 0.10, falling below the no information rate. GNB had the highest accuracy scores (0.81 $\pm$ 0.03) and high F1 score (0.83$\pm$ 0.03) creating the best overall division of the two classes. K-NN had the highest specificity followed by RF (0.80 $\pm$ 0.06), indicating that these methods are the best at identifying OSA patients (having low false positives). In addition, k-NN also had the highest PPV (0.89$\pm$0.03), while RFs had the highest NPV (0.80$\pm$0.06) score. We further note that the TAN method produced the highest sensitivity (0.91$\pm$0.05).

\begin{table}[h]
\centering
\begin{tabular}{c}
    \textbf{Performance Measures of Classification Methods on Survey Data}  \\
\end{tabular}
\begin{tabular}{p{1.6cm}p{1.5cm}p{1.5cm}p{1.6cm}p{1.6cm}p{1.6cm}p{1.4cm}}
\noalign{\smallskip}\hline \noalign{\smallskip}
Method &   Accuracy     & PPV       & NPV       & Sensitivity & Specificity
& F1 score\\ \noalign{\smallskip}\svhline \noalign{\smallskip}
LDA    & 0.61 $\pm$ 0.09 & 0.49 $\pm$ 0.10 & 0.71 $\pm$ 0.12 & 0.58 $\pm$ 0.15 & 0.64 $\pm$ 0.08 & 0.52 $\pm$ 0.11  \\
LR     & 0.54 $\pm$ 0.10 & 0.43 $\pm$ 0.12 & 0.66 $\pm$ 0.13  & 0.58 $\pm $ 0.12 & 0.52 $\pm$ 0.16 & 0.48 $\pm$ 0.10\\
DT     & 0.69 $\pm$ 0.05 &  0.77 $\pm$ 0.07  & 0.59 $\pm$ 0.14 & 0.73 $\pm$ 0.11 &  0.63 $\pm$ 0.11   & 0.58 $\pm$ 0.10       \\
RF     & 0.77 $\pm$ 0.04 & 0.70 $\pm$ 0.15 & \bf{0.80 $\pm$ 0.06} & 0.65 $\pm$ 0.10 & 0.84 $\pm$ 0.05 & 0.67 $\pm$ 0.11\\
NNET   & 0.72 $\pm$ 0.04 & 0.63 $\pm$ 0.12 & 0.79 $\pm$ 0.07 & 0.68 $\pm$ 0.12 & 0.75 $\pm$ 0.11 & 0.64 $\pm$ 0.07\\
SVM    & 0.77 $\pm$ 0.04 & 0.81 $\pm$ 0.06 & 0.69 $\pm$ 0.13  & 0.81 $\pm$ 0.07 & 0.69 $\pm$ 0.07 & 0.69 $\pm$ 0.08 \\
k-NN   & 0.74 $\pm$ 0.07  &  \bf{0.89 $\pm$ 0.03} & 0.61 $\pm 0.13$ & 0.67 $\pm$ 0.10 & \textbf{0.86 $\pm$ 0.05}  & 0.71 $\pm$ 0.11 \\
GNB     & \bf{0.81 $\pm$ 0.03} &    0.88 $\pm$ 0.03            & 0.70$\pm$ 0.10  & 0.79 $\pm$ 0.06          & 0.81 $\pm$ 0.07   & \bf{0.83 $\pm$ 0.03}     \\
TAN    & 0.78 $\pm$ 0.05 & 0.77 $\pm$ 0.07 & 0.79 $\pm$ 0.12   & \bf{0.91 $\pm$ 0.05}  & 0.56 $\pm$ 0.10 & 0.83 $\pm$ 0.04                      \\
SHNB   & 0.75 $\pm$ 0.03  & 0.78 $\pm$ 0.06           & 0.70 $\pm$ 0.16          & 0.85 $\pm$ 0.07             & 0.57 $\pm$ 0.12 & 0.81$\pm$ 0.05\\
\noalign{\smallskip}\svhline \noalign{\smallskip}
Spectral             &0.55       &0.63    &0.42    &0.62      &0.43 & 0.63\\
Threshold $q_k$      & \bf{0.77}       & \bf{0.79}    & \bf{0.75}       & \bf{0.87}      & \bf{0.63} & \bf{0.83}\\
\noalign{\smallskip}\hline \noalign{\smallskip}
\end{tabular}
\caption{Performance measures for all supervised and unsupervised learning methods applied to the survey data set. Best performances by each metric are bolded. Supervised methods were applied to ten different train/test splits; their performance measures are recorded in the format of mean performance plus/minus standard deviation. Unsupervised learning methods do not take training sets, so only mean performance is reported. Note that the Cut-Cluster-Classify and DBSCAN methods classified all data as having OSA, while the CkNN method classified all but one subject as having OSA; see Figure \ref{fig:dbc_results}. We do not consider these methods to be informative.}
\label{tab:survey_only}
\end{table}

\begin{table}[h]
\centering
\begin{tabular}{c}
    \textbf{Adjusted Rand Index for clustering methods on Survey Data}  \\
\end{tabular}
\begin{tabular}{p{1.6cm}p{1.5cm}p{1.5cm}p{1.6cm}p{1.6cm}p{1.6cm}p{1.4cm}}
\noalign{\smallskip}\hline \noalign{\smallskip}
 &   DBSCAN     & CCC       & Spectral      & CkNN & Threshold $q_k$ \\
 
 \noalign{\smallskip}\svhline \noalign{\smallskip}

ARI     & 0 &  0 &  0.0023 & 0.0067 & \textbf{0.2937}
\\
\noalign{\smallskip}\hline \noalign{\smallskip}
\end{tabular}
\caption{Adjusted Rand index presented for clustering methods on survey data. Values close to 0 indicate a random chance results while those closer to 1 indicate a perfect clustering. Note that DBSCAN and CCC produced only one cluster.}
\label{tab:ARI_Survey}
\end{table}

Amongst the clustering methods, DBSCAN, CCC classified all, and CkNN classified almost all data into one cluster thereby giving no information - the ARI scores for these methods are very close to 0. On the other hand, spectral clustering produced two clusters but had an accuracy of only 55\%, below our no-information rate of 58\%. The ARI of this method (0.0023) falls below that of CkNN (0.0067) indicating its poor performance.  Threshold $q_k$ density produces the best results amongst clustering methods with an accuracy of 0.77, sensitivity of 0.87 and an F1 score of 0.83, the latter two being competitive with supervised methods. It lacked the most in specificity (0.62). The ARI of this method was 0.2937 - indicating a fair but not ideal clustering. 
All performance values for techniques as applied to survey data can be found in Table \ref{tab:survey_only} and \ref{tab:ARI_Survey}.

\subsection{Results for Craniofacial Data}

\paragraph{Results: CF Distributions} In comparing the distributions of craniofacial variables, we found that the three variables with the largest difference between patients and controls were Palate Score, Lower Face Height, and Overjet Score. The other variables may have some difference in distribution shapes (see Table \ref{tab:CF_summary} and \textbf{Figures \ref{fig:CF_graphs} and \ref{fig:DTS})}, but overall do not appear as different when comparing control subjects and OSA subjects. This is not to say that all other craniofacial variables should be ignored; rather, clinicians should pay extra attention to variables Palate Score, Lower Face Height, and Overjet Score when evaluating a patient.

\paragraph{Results: Classification with Craniofacial Data} 

Table \ref{tab:CF1} displays results of various classification methods applied to craniofacial data. Overall, unlike survey data, there was no one method that stood out over the rest in the overall performance, with multiple methods achieving the highest value for each measure.


\begin{table}[h]
	\centering
	\begin{tabular}{c}
		\textbf{Performance Measures of Classification Methods on Craniofacial Data}
	\end{tabular}
	\begin{tabular}{p{1.6cm}p{1.5cm}p{1.5cm}p{1.6cm}p{1.6cm}p{1.6cm}p{1.4cm}}
		\noalign{\smallskip}\hline\noalign{\smallskip}
		Method &   Accuracy     & PPV       & NPV       & Sensitivity & Specificity & F1 score\\ 
		\noalign{\smallskip}\svhline\noalign{\smallskip}
		LDA     & 0.63 $\pm$ 0.07 & 0.49 $\pm$ 0.22 &   0.74 $\pm$ 0.11 & 0.59 $\pm$ 0.26 & 0.70 $\pm$ 0.18 &  0.56 $\pm$ 0.09\\
		LR      & 0.66 $\pm$ 0.06 & 0.59 $\pm$ 0.18 & 0.76 $\pm$ 0.09 & 0.63 $\pm$ 0.16 & \textbf{0.71 $\pm$ 0.16} & 0.57 $\pm$ 0.08\\
		DT     & \textbf{0.70 $\pm$ 0.06} & 0.78 $\pm$ 0.08 & 0.62 $\pm$ 0.16 & 0.76 $\pm$ 0.11 &  0.63 $\pm$ 0.16  & 0.60 $\pm$ 0.11\\
		RF      & 0.66 $\pm $ 0.07 & 0.56 $\pm $ 0.15 & \bf{0.77 $\pm$ 0.09} & 0.67 $\pm$ 0.15 & 0.67 $\pm$ 0.12 & 0.59 $\pm$ 0.11 \\
		NNET    & 0.67 $\pm$ 0.08  & 0.58 $\pm$ 0.17 & \bf{0.77 $\pm$ 0.09} & 0.67 $\pm$ 0.13 & 0.69 $\pm$ 0.15 & 0.60 $\pm$ 0.09\\
		SVM    & \textbf{0.70 $\pm$ 0.06} &  0.77 $\pm$ 0.08 &  0.62 $\pm$ 0.17 & \textbf{0.76 $\pm$ 0.10}  & 0.63 $\pm$ 0.14 & 0.60 $\pm$ 0.10 \\
		k-NN   & 0.66 $\pm$ 0.07  & 0.77 $\pm$ 0.11 & 0.56 $\pm$ 0.17  & 0.67 $\pm$ 0.14 & 0.67 $\pm$ 0.18 & 0.58 $\pm$ 0.11 \\
		GNB      &   0.69 $\pm$ 0.05 & \bf{0.78 $\pm$ 0.07}      & 0.59 $\pm$ 0.13          & 0.71 $\pm$ 0.09            & 0.67 $\pm$ 0.09 & \bf{ 0.74 $\pm$ 0.04}          \\
		TAN  &  0.67 $\pm$ 0.07             & 0.77 $\pm$ 0.07               & 0.56 $\pm$ 0.16         &  0.70 $\pm$ 0.11         &    0.64 $\pm$ 0.14 & 0.72$\pm$ 0.05                     \\
		SHNB    & {0.69 $\pm$ 0.05}               &     0.78 $\pm$ 0.08      &  0.58 $\pm$ 0.13         &           0.72 $\pm$ 0.08  & 0.66 $\pm$ 0.12   & \bf{0.74 $\pm$ 0.05}          \\
		\noalign{\smallskip}\svhline \noalign{\smallskip}
		Spectral   &0.53         &0.65     &0.42    &0.51      & \bf{0.57} & 0.57\\
		CkNN    & \bf{0.71}         &\bf{0.71}    & \bf{0.69}    &\bf{0.88}     &0.43 & \bf{0.78}\\
		Threshold $q_k$     & \bf{0.71} & \bf{0.71}    & \bf{0.69}    &\bf{0.88}     &0.43 & \bf{0.78} \\ \hline
	\end{tabular}
	\caption{Performance measures for all supervised and unsupervised learning methods applied to the craniofacial data set. For each metric, the best performance, as considered both by mean and standard deviation, is in bold. Supervised methods were applied to ten different train/test splits; their performance measures are recorded in the format of mean performance plus/minus standard deviation. Unsupervised learning methods do not take training sets, so only mean performance in classification is reported. Note that the Cut-Cluster-Classify method and DBSCAN classified almost all data as having OSA. As such, we do not consider these methods to be informative.}
	\label{tab:CF1}
\end{table}


Support vector machines and DT had the best accuracy (0.70 $\pm$ 0.06), GNB, SHNB and DT produced the best PPV (0.78 $\pm$ 0.07, 0.78 $\pm$ 0.08, 0.78 $\pm$ 0.08), NNET and RF tied for best NPV (0.77 $\pm$ 0.09), SVM and DT shared the best sensitivity (0.76 $\pm$ 0.10, 0.76 $\pm$ 0.11), and LR had the best specificity (0.71 $\pm$ 0.16). Note that GNB and SHNB had the highest F1 score (0.74 $\pm$ 0.04 and 0.74 $\pm$ 0.05 respectively), with a competitive accuracy, producing the best division of the two classes amongst the methods explored.

It follows that if we only use craniofacial data, DT and SVM have the lowest false negative rates helping identify patients with no OSA, while LR, surprisingly, produces the lowest false positives (followed closely by LDA and NNET) whence being ideal for identifying subjects who do have OSA. Compared to the best performances in these measures in survey data, the means of measures are lower and the standard deviations are larger, indicating that the  performance is worse and less stable. 

\begin{table}[h]
\centering
\begin{tabular}{c}
    \textbf{Adjusted Rand Index for clustering methods on Survey Data}  \\
\end{tabular}
\begin{tabular}{p{1.6cm}p{1.5cm}p{1.5cm}p{1.6cm}p{1.6cm}p{1.6cm}p{1.4cm}}
\noalign{\smallskip}\hline \noalign{\smallskip}
 &   DBSCAN     & CCC       & Spectral      & CkNN & Threshold $q_k$ \\
 
 \noalign{\smallskip}\svhline \noalign{\smallskip}

ARI     & 0 &  -0.0042 & -0.0016  & \textbf{0.1536} & \textbf{0.1536}
\\
\noalign{\smallskip}\hline \noalign{\smallskip}
\end{tabular}
\caption{Adjusted Random index presented for clustering methods on craniofacial data. Values close to 0 indicate a random chance results while those closer to 1 indicate a perfect clustering. Negative values indicate performance worse than random chance.}
\label{tab:ARI_Cranio}
\end{table}
For the unsupervised methods, CCC and DBSCAN classifies almost all data into one cluster, thereby producing almost no information and achieving ARIs close to 0. Spectral clustering continues to perform poorly with an accuracy of 53\% falling below the no information rate of 58\%. CkNN and Threshold $q_k$ performed the best with identical results. Both had an accuracy of 0.71\%, a sensitivity of 0.88\% and a F1 score of 78\% beating the supervised methods. These scores indicate that out of all methods, CkNN and Threshold $q_k$ are best at dividing the patients into the two classes overall and can help exclude OSA in subjects that do not have it. However, a low ARI score of 0.1536 indicates that the results are not much better than random clustering. Further, these methods suffer a low specificity of 0.43. 

We can conclude that while the craniofacial data can aid in diagnosis, it cannot be used alone. On the one hand, the poor results of classification in craniofacial data may be a self-fulfilling prophesy: OSA diagnosis in the clinical studies is based on both, the questions and the craniofacial measurements. Therefore, there exist patients who have normal craniofacial measurements but were still diagnosed with OSA. On the other hand, the existence of OSA patients and controls having the exact same craniofacial measurements indicates  that craniofacial data by itself cannot be used to distinguish the two groups. Additionally, unsupervised methods seem to produce the best results on this data, granted the clusters are labelled in an optimal manner.


\subsection{Results for Combined Survey and Craniofacial Data}
Lastly, we analyze the combined survey and craniofacial data to see how the two work together in classification. Overall, the performance of each method was similar to each method's respective performance for survey data. Supervised algorithms applied to the combined data generally outperformed algorithms applied only to survey data or craniofacial data (see Tables \ref{tab:combo}, \ref{tab:CF1} and \ref{tab:survey_only}).

\begin{table}[]
\centering
\begin{tabular}{c}
     \textbf{Performance Measures of Classification Methods on Combined Data}  \\
\end{tabular}
\begin{tabular}{p{1.6cm}p{1.5cm}p{1.5cm}p{1.6cm}p{1.6cm}p{1.6cm}p{1.4cm}}
\noalign{\smallskip}\hline\noalign{\smallskip}
Method  &   Accuracy& PPV  & NPV  & Sensitivity & Specificity & F1 score \\ 
\noalign{\smallskip}\svhline\noalign{\smallskip}
LDA    & 0.65 $\pm$ 0.09   & 0.53 $\pm$ 0.12   & 0.77 $\pm$ 0.12  & 0.70 $\pm$ 0.15 &  0.63 $\pm$ 0.09   & 0.59 $\pm$ 0.10\\
LR   &  0.52 $\pm$ 0.09  & 0.41 $\pm$ 0.11 & 0.64 $\pm$ 0.12 & 0.56 $\pm$ 0.11 & 0.50 $\pm$ 0.15 & 0.46 $\pm$ 0.09\\
DT     & 0.69 $\pm$ 0.05 &  0.76 $\pm$ 0.06  & 0.58 $\pm$ 0.14 & 0.73 $\pm$ 0.08 &  0.63 $\pm$ 0.10 &    0.56 $\pm$ 0.12      \\
RF     & 0.78 $\pm$ 0.03  & 0.72 $\pm$ 0.10 & \bf{0.81 $\pm$ 0.06} & 0.67 $\pm$ 0.08 & 0.85 $\pm$ 0.04  & 0.69 $\pm$ 0.07\\
NNET   & 0.75 $\pm$ 0.04 & 0.67 $\pm$ 0.12  & 0.80 $\pm$ 0.08  &  0.69 $\pm$ 0.11   & 0.80 $\pm$ 0.07 & 0.67 $\pm$ 0.08\\
SVM    & 0.76 $\pm$ 0.06  & 0.81 $\pm$ 0.06  & 0.70 $\pm$ 0.16 & 0.81 $\pm$ 0.11 &  0.68 $\pm$ 0.11  & 0.67 $\pm$ 0.10 \\
k-NN   &  0.75 $\pm$ 0.08  & \bf{0.89 $\pm$ 0.04}& 0.62 $\pm$ 0.15  &  0.69 $\pm$ 0.10 & \bf{0.86 $\pm$ 0.07}     & 0.72 $\pm$ 0.12 \\
GNB     &\bf{0.80 $\pm$ 0.03}                &         0.88 $\pm$ 0.03  & 0.70 $\pm$ 0.10          &       {0.79$\pm$ 0.06}      &  0.81 $\pm$ 0.07        & \bf{0.83$\pm$ 0.03}     \\
TAN    & 0.77 $\pm$ 0.04               &         0.76 $\pm$ 0.07 & 0.79 $\pm$ 0.12 & \bf{0.91 $\pm$ 0.04}           &  0.55 $\pm$ 0.09  &       \bf{0.83 $\pm$ 0.03}               \\
SHNB   &  0.76 $\pm$ 0.03              &         0.77 $\pm$ 0.08  & 0.72 $\pm$ 0.13           &      0.87 $\pm$ 0.06       & 0.57 $\pm$ 0.13   &   0.81$\pm$ 0.02      \\
\noalign{\smallskip}\svhline \noalign{\smallskip}
Spectral &0.51         &0.60    &0.38      &0.58      &0.40 & 0.59\\

Threshold $q_k$   & \bf{0.78}        & \bf{0.79}    & \bf{0.76}    &\bf{0.88}     & \bf{0.63} & \textbf{0.83}\\
\hline

\end{tabular}
\caption{Performance measures for all supervised and unsupervised learning methods applied to the combined data set. Best performances by each metric are bolded. Supervised methods were applied to ten different train/test splits; their performance measures are recorded in the format of mean performance plus/minus standard deviation. Unsupervised learning methods do not take training sets, so there is only one number reported for their performance measures. Note that the Cut-Cluster-Classify method classified all data as having OSA, while the CkNN and DBSCAN methods classified almost all data as having OSA; see Figure \ref{fig:dbc_results}. As such, we do not consider these methods to be informative.}
\label{tab:combo}
\end{table}

 Once again, GNB had the highest accuracy (0.80 $\pm$ 0.03) and F1 score (0.83$\pm$ 0.03) and a competitive PPV (0.88 $\pm$ 0.03) indicating a good overall division of classes. TAN matched this F1 score and had the highest sensitivity (0.91 $\pm$ 0.04), thereby being the best method for excluding OSA in subjects (having a low false negative rate).   The highest PPV (0.89$ \pm$ 0.04) and specificity (0.86$ \pm$ 0.07) was achieved by k-NN, making it the best method for ensuring a low false positive rate rate, i.e. for identifying OSA patients. RF closely follows in specificity (0.85 $\pm$ 0.04) and also achieves the highest NPV (0.81 $\pm$ 0.06).   All the other metrics for performance either stayed about the same or slightly improved.

\begin{figure}
    \centering
    \includegraphics[height=6in]{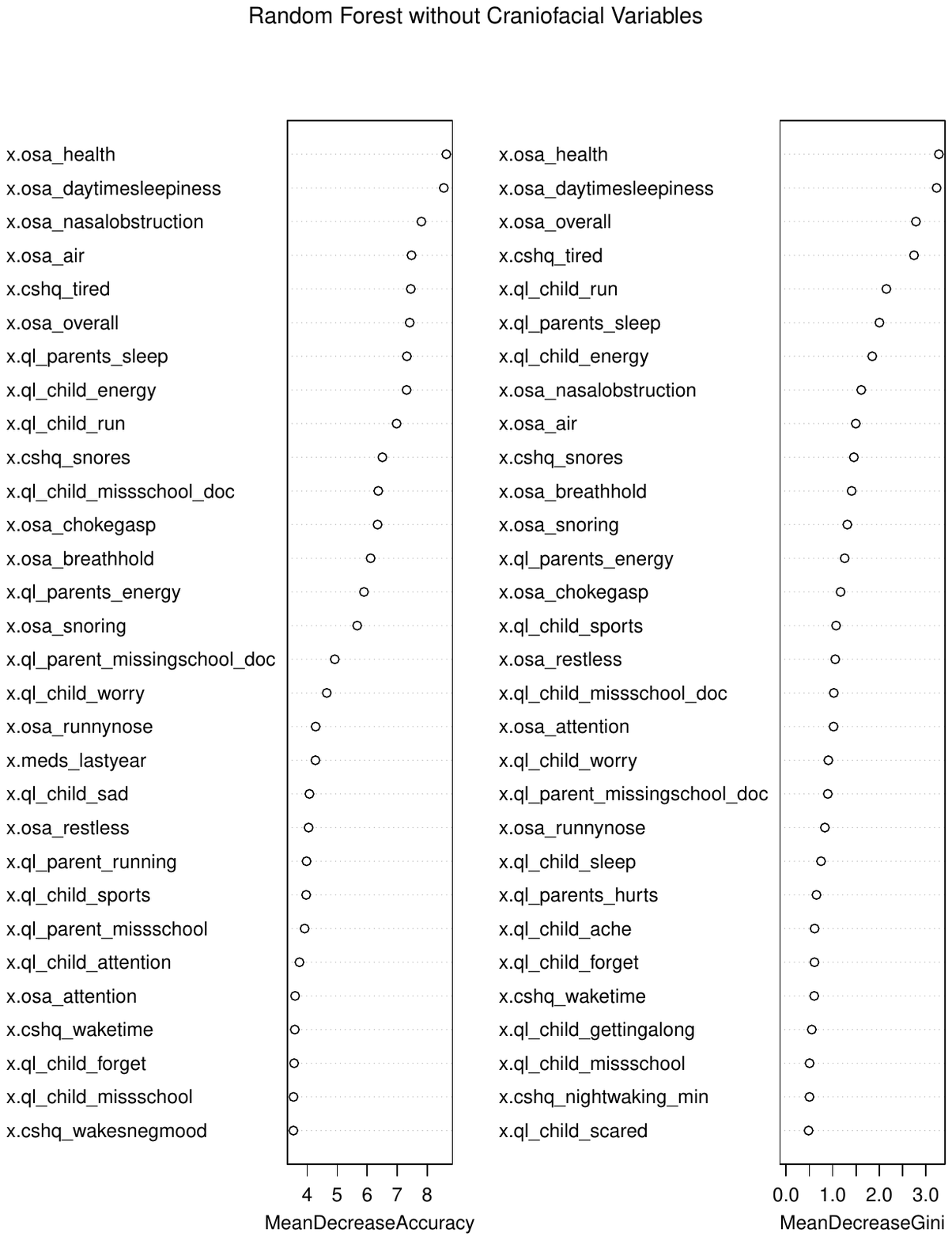} 
    \caption{The plot of variable importance from RF for the survey data. The left plot ranks variables by their contributions to the accuracy, while the right ranks variables for their decrease in Gini score. The top three most important variables were from the same OSA-18 questionnaire, but questions from each of the three surveys made the list.}
    \label{fig:varImportantNO}
\end{figure}

 \textbf{Figures \ref{fig:varImportant} and \ref{fig:varImportantNO}} shows the important variables identified by RF for the combined and Survey data sets respectively. There is only one craniofacial variable identified as an important variable, i.e., \textit{``palate score''}. This information is especially valuable given that random forests performs competitively in many measures.

\begin{table}[h]
\centering
\begin{tabular}{c}
    \textbf{Adjusted Rand Index for clustering methods on combined Data}  \\
\end{tabular}
\begin{tabular}{p{1.6cm}p{1.5cm}p{1.5cm}p{1.6cm}p{1.6cm}p{1.6cm}p{1.4cm}}
\noalign{\smallskip}\hline \noalign{\smallskip}
 &   DBSCAN     & CCC       & Spectral      & CkNN & Threshold $q_k$ \\
 
 \noalign{\smallskip}\svhline \noalign{\smallskip}

ARI     & 0.0137 &  0 & -0.0065 & 0.0067  & \textbf{0.3064}
\\
\noalign{\smallskip}\hline \noalign{\smallskip}
\end{tabular}
\caption{Adjusted Random index presented for clustering methods on combined survey and craniofacial data. Values close to 0 indicate a random chance results while those closer to 1 indicate a perfect clustering. Negative values indicate performance worse than random chance. Note that CCC produced only one cluster.}
\label{tab:ARI_Combined}
\end{table}

For unsupervised methods, CCC produced a single cluster, similarly to the Survey data, while CkNN and DBSCAN classified almost all data into the OSA cluster. These methods had ARI values close to 0 as displayed in Table \ref{tab:ARI_Combined}, thereby not producing very informative results. Spectral clustering continued to perform poorly with a low accuracy of 51\% and the worst ARI score of -0.0065. Again, threshold $q_k$ performed best with a competitive accuracy and F1 score of 0.78 and 0.83 respectively, and a sensitivity of 0.88. As in the previous results, this would be good for excluding OSA in patients. We note that it gives the highest ARI score thus far at 0.3064 out of all datasets and clustering methods considered. \textbf{Figure \ref{fig:dbc_results}} further helps illustrate that clustering by a threshold on density gives more meaningful labels than the more complex methods. We also note that other methods tend to produce only one cluster or very spread out clusters.

From the results, we observe that incorporating craniofacial data and the survey data together do give the best performance in just about every metric. While craniofacial data does not perform well on its own, it does contribute to OSA classification when combined with the survey data. 

\section{Conclusion and Future Research}\label{sec:conclusion}
We carried out Friedman's test~\cite{Demsar} for each performance measure using Survey and Craniofacial Index  (Tables~\ref{tab:survey_only} and \ref{tab:CF1}). For the supervised learning methods~(classification), PPV~(p-value $=0.017$),  sensitivity~(p-value $=0.005$), and F1-score~(p-value $=0.005$) indicate statistical significance  However, the Nemenyi test of the pairwise comparison was not powerful enough to detect any significant differences between the 10 classification methods. We did not perform Friedman's test for unsupervised (clustering) methods due to few number of methods.

\begin{figure}
    \centering
    \includegraphics[height=6in]{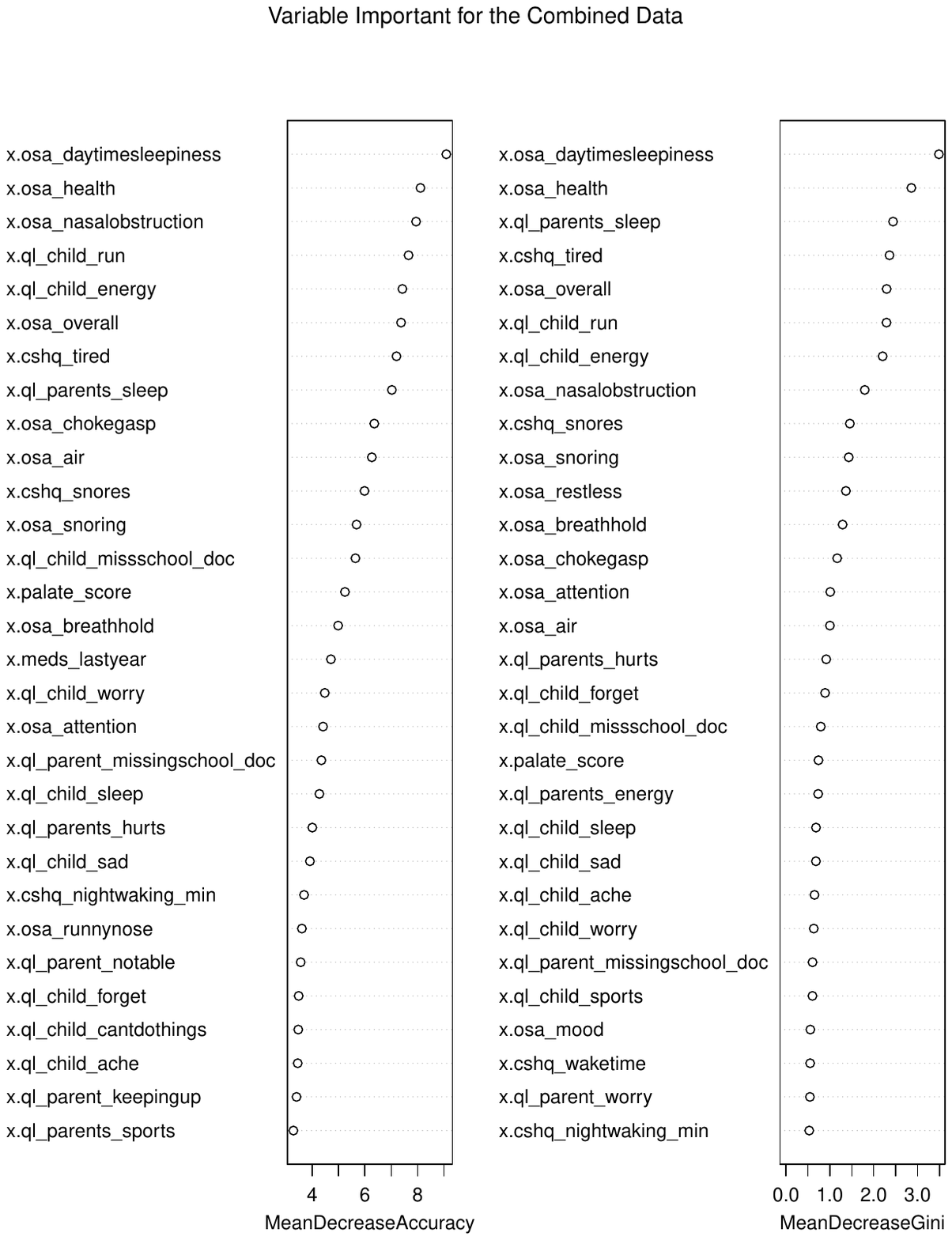} 
    \caption{The plot of variable importance from random forests for the combined data. The left plot ranks variables by their contributions to the mean decrease of the accuracy, while the right ranks variables for their decrease in Gini score. The only craniofacial variable marked as important was \textquotedblleft palate score"; all other highly ranked variables were from the surveys.}
    \label{fig:varImportant}
\end{figure}

In addition to assessing methods of classification, by considering three subsets of the data (survey data only, craniofacial data only, and both combined) and using interpretable classification methods such as random forests, we were able to look at which variables played the largest role in separating OSA patients from controls. Survey data usually had results very similar to that of combined data, likely because there were only 8 craniofacial measurements considered in our data set. Although performance on the survey data seem to be better than that on the craniofacial data, in Section 3, craniofacial data was able to detect some OSA cases in the K-Mapper algorithm that the survey data and combined data could not (see \textbf{Figure} \ref{fig:KMapper_cheap}). This suggests that it is worth considering the two data sets both, separately and combined, in future algorithms. With the random forest algorithm, as shown in \textbf{Figure} \ref{fig:varImportant}, various survey questions such as those addressing daytime sleepiness and overall health seem more important than the craniofacial variables, with the exception of palate score. Overall, the survey data performed better in binary classification than the craniofacial data. However, the latter did add useful information based on the slight improvements in performance going from just survey data to combined data (compare Tables 	\ref{tab:survey_only} and \ref{tab:combo}).

Our results demonstrate that using inexpensive data can still make strong predictions for a binary classification of being at a risk of OSA vs not having OSA. In particular, random forests, Gaussian Naive Bayes classifiers, and $k$-nearest neighbors performed the best. Of those successful methods, Random Forests and Gaussian Naive Bayes are interpretable in that we can see which variables were prioritized most in sorting. This information is valuable for clinicians in obtaining a differential diagnosis.

\begin{figure}
    \centering
    \begin{tabular}{cc}
     \includegraphics[width=2.2in]{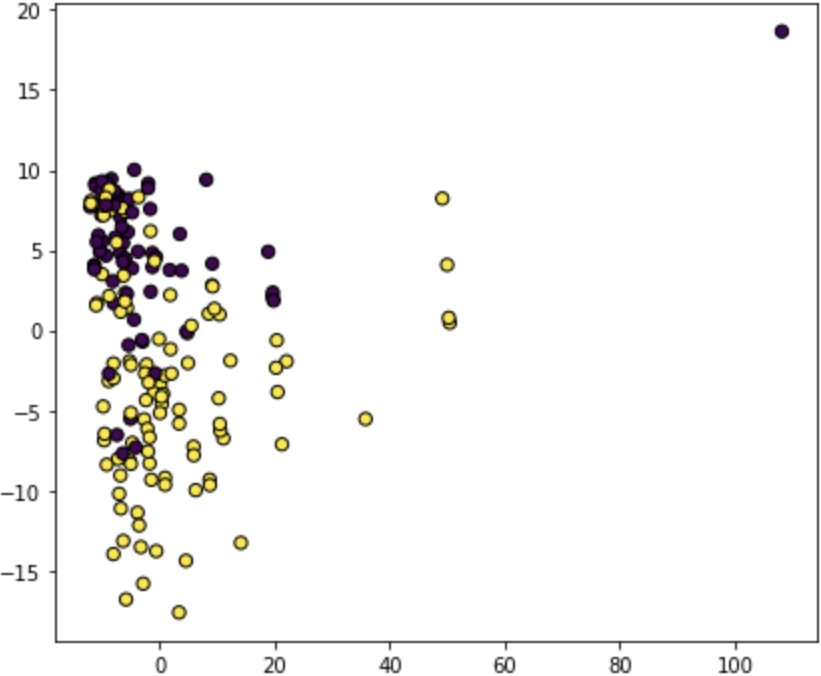}&  
     \includegraphics[width=2.2in]{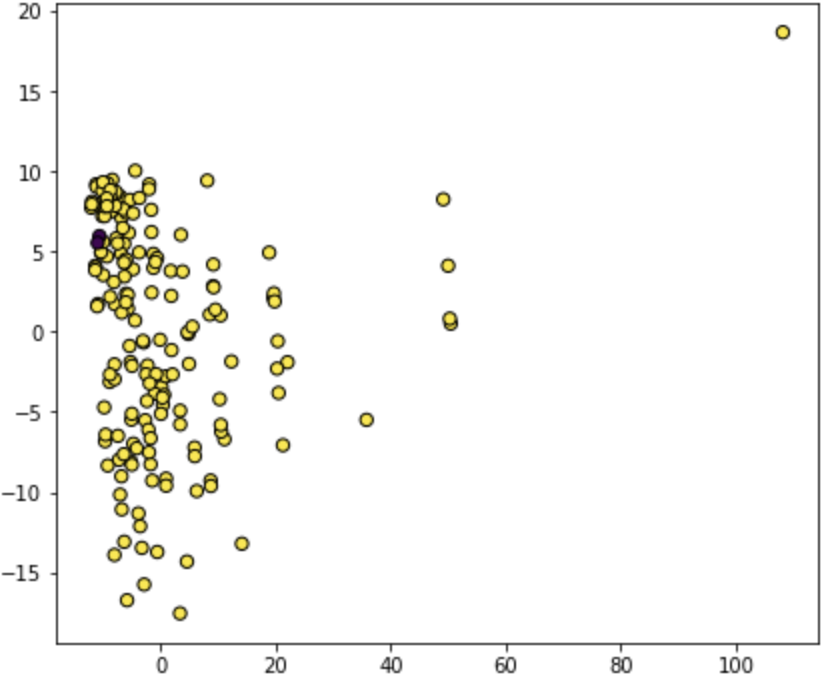}\\
      \scriptsize{Ground truth}   & \scriptsize{DBSCAN} \\
     \includegraphics[width=2.2in]{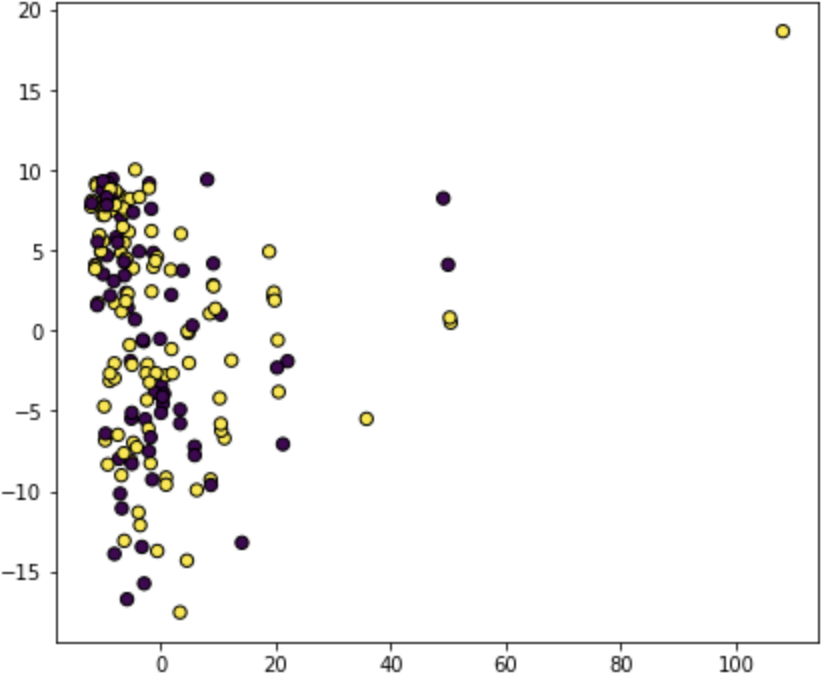}&
     \includegraphics[width=2.2in]{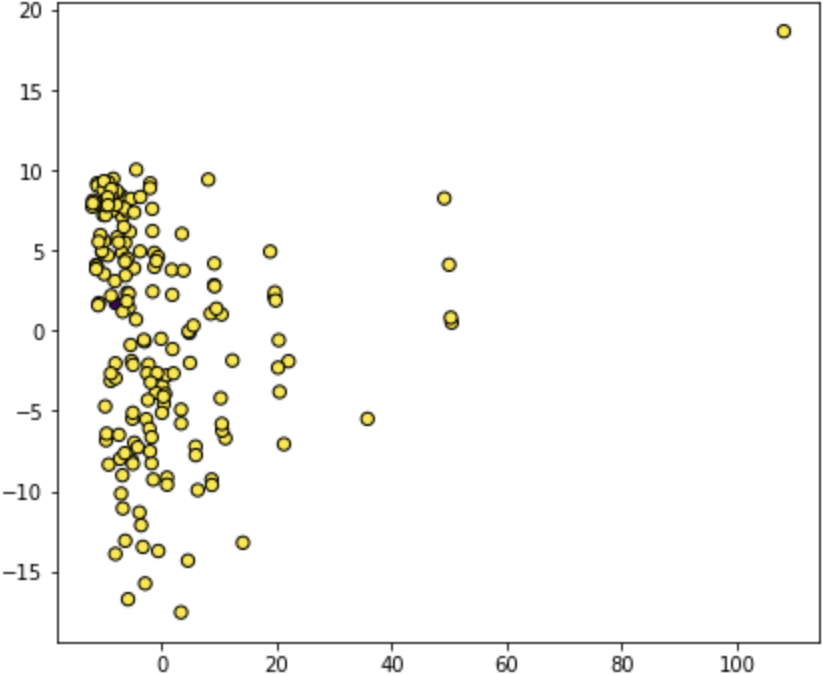}\\
     \scriptsize{Spectral clustering} & \scriptsize{CkNN} \\
     \includegraphics[width=2.2in]{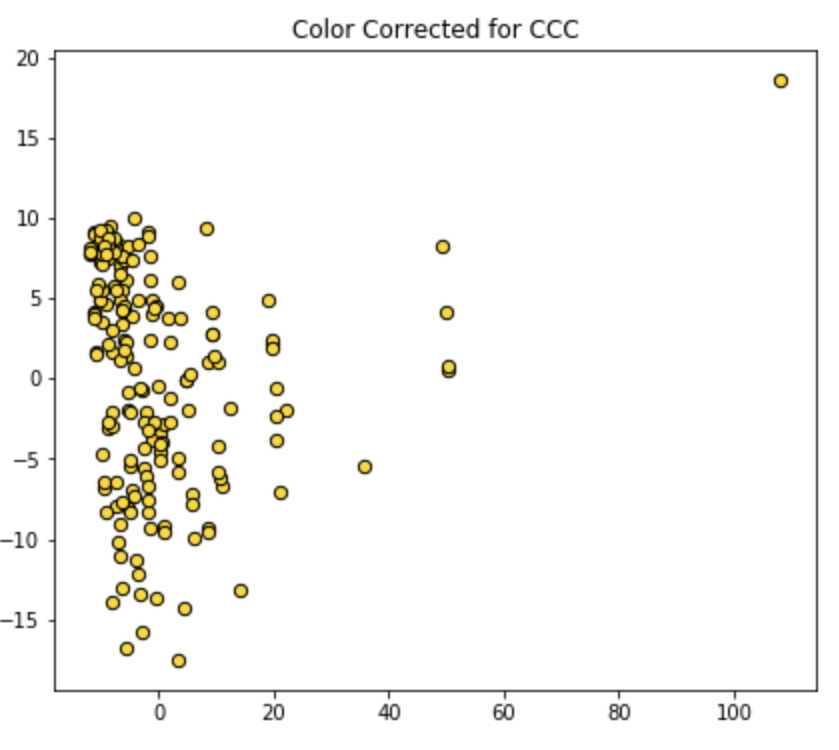}& 
     \includegraphics[width=2.2in, height=1.8in]{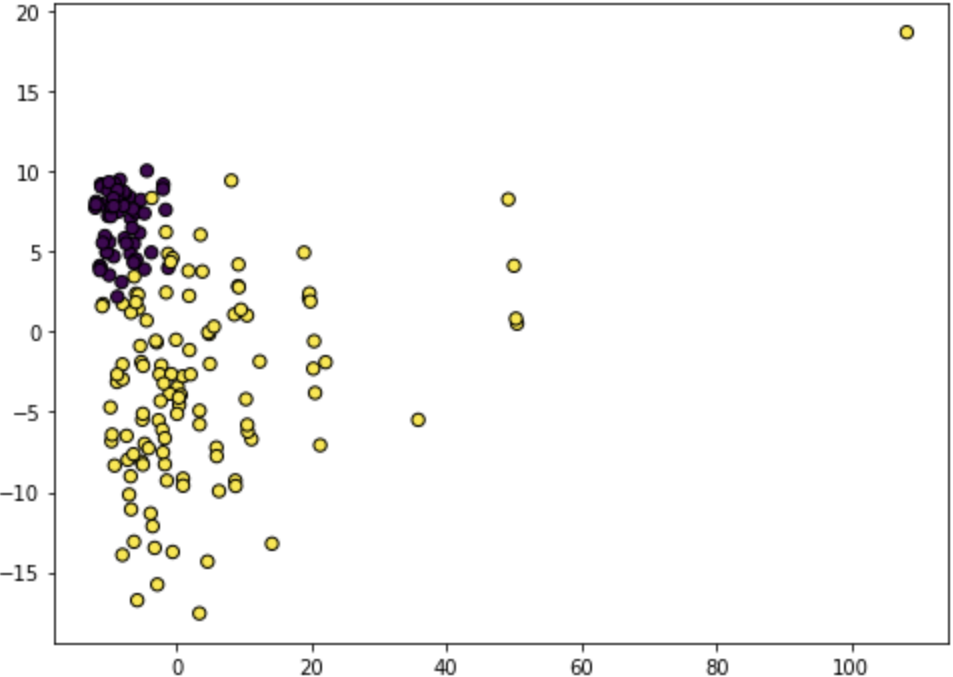}\\
     \scriptsize{Cut-Cluster-Classify (CCC)} & \scriptsize{Threshold sample density}\\
    \end{tabular}
    \caption{Kernel PCA coordinates of the combined survey and craniofacial data colored by the resulting labels from each clustering method starting with the ground truth. In this plot, blue (dark) points correspond to the control group while the yellow (light) ones correspond to OSA patients.}\label{fig:dbc_results}
\end{figure}

Unsupervised (clustering) methods often clustered all or most data into one class, providing uninformative results. Threshold $q_k$ was an exception, with a consistent performance. Additionally, CkNN and threshold $q_k$ (in the best case scenario) seem to perform better on craniofacial data on several metrics compared to supervised learning.
As shown in \textbf{Figure \ref{fig:dbc_results}}, the ground truth plot conjectures the hypothesis that points in the control group tend to be in a high density region while those in the patient group tend to be spread out in the data cloud.
Using thresholding best allows exploiting the difference in densities in the two classes, while methods such as spectral clustering seemed to produce clusters with similar densities (again, see \textbf{Figure \ref{fig:dbc_results}}). 
From these results we can only conclude that more points are needed in order to apply a clustering approach and clustering techniques. Further, on our dataset, considering clustering techniques that prioritize a difference in sample density between classes would be valuable.

Using survey data and craniofacial data, we merely attempted to classify whether a subject was diagnosed with OSA or not. However, this is a simplification of the standard clinical classification of patients into the categories of having as no, mild, moderate, or severe OSA, based on the apnea hypopnea index. The next natural step in our work is to expand classification techniques to these four possible outcomes and identifying which predictors indicate OSA severity. It would be interesting to explore how the spreading out of OSA patients is affected by the severity of the condition (no, mild, moderate or severe OSA).

Another interesting future direction is to repeat the algorithms but instead of combining multiple surveys, isolate each survey and run each classification method separately. In doing so, we may lend evidence to which survey is the most effective in diagnosing OSA. We would do this for both the binary OSA classification and the classification by the apnea-hypopnea index. 

We hope to use this information to obtain an algorithm which can aid clinicians in diagnosis and personalized treatment of a child with OSA. This algorithm would be updated as we track children through treatment and follow up in their progress towards healthy sleeping patterns.

\begin{acknowledgement}
We thank The Institute for Computational and Experimental Research in Mathematics~(ICERM), Brown University for hosting the second Women in Data Science and Mathematics workshop (WiSDM 2) in summer 2019. We would also like to thank the other group members, Brenda Praggastis, Kritika Singhal, Melissa Stockman, and Sarah Tymochko. Finally, we would like to thank the reviewers for their constructive feedback and their help improving the manuscript.
EW is funded by the National Science Foundation Graduate Research Fellowship Program under Grant No. 1644760. Any opinions, findings, and conclusions or recommendations expressed in this material are those of the author(s) and do not necessarily reflect the views of the National Science Foundation. XW would like to thank the National Sciences and Engineering Research Council of Canada (NSERC DG 2019 - 05917). GH would like to thank the National Sciences and Engineering Research Council of Canada (NSERC DG 2016-05167),
Seed grant from Women and Children's Health Research Institute, Biomedical Research Award from American Association of Orthodontists Foundation, and the McIntyre Memorial fund from the School of Dentistry at the University of Alberta. 
\end{acknowledgement}

\newpage
\section{Appendix}
%

\begin{table}
    \centering
    \begin{tabular}{c}
         \textbf{Frequencies and Earth Mover's Distance for Craniofacial Data Measures}  \\
    \end{tabular}
    \begin{tabular}{p{3cm}p{1.5cm}p{1.5cm}p{1.5cm}p{1.5cm}p{1.5cm}}
        \noalign{\smallskip}\hline\noalign{\smallskip}
         Metric & Group & Score 0 & Score 1 & Score 2 & EMD \\ 
         \noalign{\smallskip}\svhline\noalign{\smallskip}
         Profile & P & 0.7477 & 0.0 & 0.2522 & 0.1243 \\ 
          & C & 0.9342 & 0.0 & 0.0658 &  \\ 
         Midface Deficiency & P & 0.6126 & 0.3513 & 0.0360 & 0.1179 \\ 
          & C & 0.7895 & 0.1934 & 0.0132 &  \\ 
         Lower Face Height & P & 0.5766 & 0.3604 & 0.0630 & 0.1507 \\ 
          & C & 0.8026 & 0.1579 & 0.0395 &  \\ 
          Lip Strain & P & 0.6306 & 0.2793 & 0.0900 & 0.1234  \\ 
          & C & 0.8158 & 0.1711 & 0.0132 &  \\ 
          Palate & P & 0.4775 & 0.4595 & 0.0630 & 0.1817 \\ 
          & C & 0.7500 & 0.2105 & 0.0395 &  \\ 
          Overjet & P & 0.6306 & 0.0 & 0.3694 & 0.1498 \\ 
          & C & 0.8553 & 0.0 & 0.1447 & \\ 
          Overbite & P & 0.8919 & 0.0 & 0.1081 & 0.0633 \\ 
          & C & 0.9868 & 0.0 & 0.0132 &  \\ 
          Posterior Bite & P & 0.8468 & 0.0811 & 0.0721 & 0.0846 \\ 
          & C & 0.9737 & 0.0132 & 0.0132 &  \\ \hline
         
    \end{tabular}
    \caption{Table showing the frequencies of scores 0, 1, and 2 for the patient (P) group and the control (C) group for each craniofacial metric, as well as the Earth Mover's Distance between these frequency distributions. According to the data, the craniofacial variables with the largest distribution differences were palate score, lower face height, and overjet. However, as shown in Figure \ref{fig:varImportant}, only the palate score played a significant role in classification in the combined data set.}
    \label{tab:CF_summary}
\end{table}

\begin{figure}
    \centering
    \begin{tabular}{cc}
     \includegraphics[width=2.2in,height=1.95in]{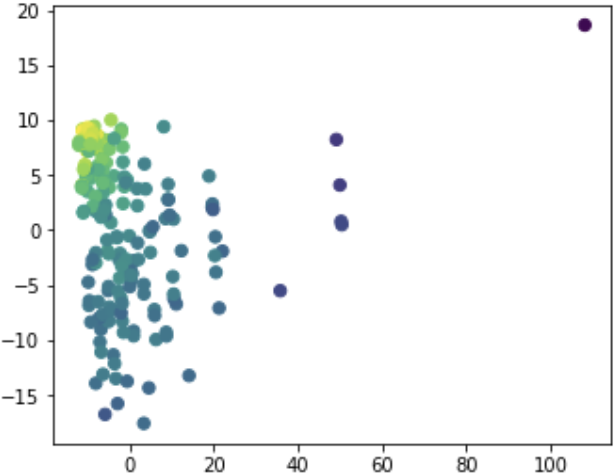}&  
     \includegraphics[width=2in]{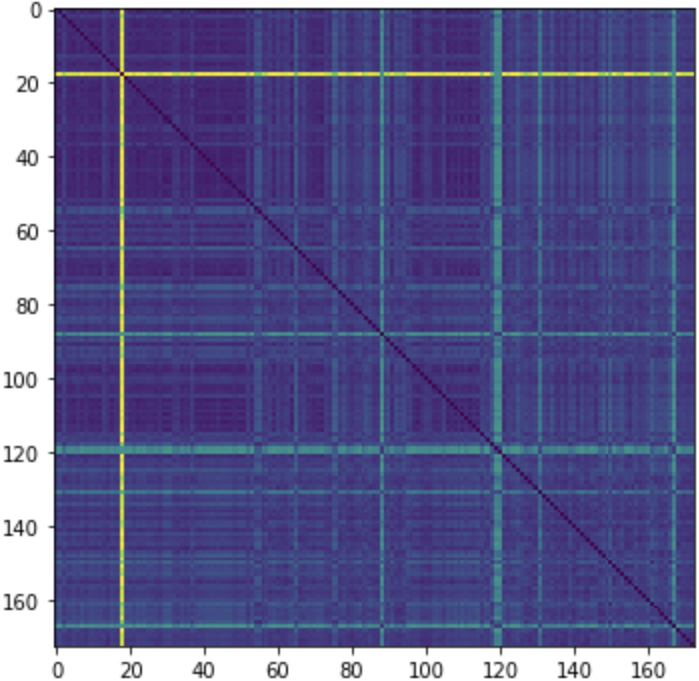}\\
      \scriptsize{(a)}   & \scriptsize{(b)} \\
     \includegraphics[width=2in]{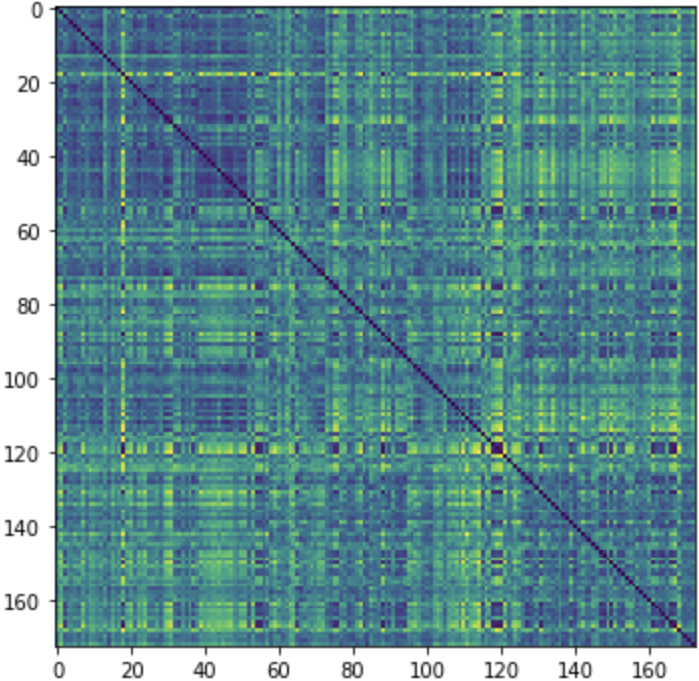}&
     \includegraphics[width=2in]{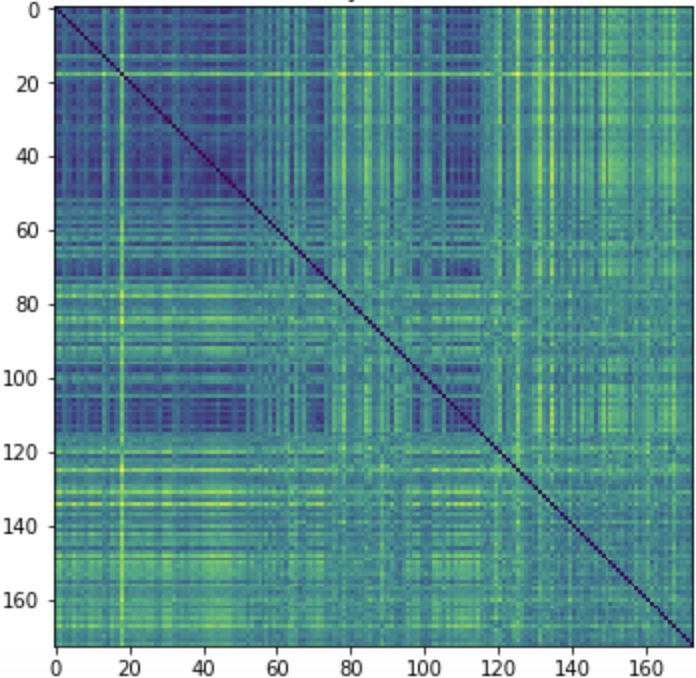}\\
     \scriptsize{(c)} & \scriptsize{(d)} 
    \end{tabular}
    \caption{(a) Kernel PCA projection of all features of the data colored by sample density and the distance matrices for the data set according to (b) Euclidean, (c) Correlation, and (d) Manhattan metrics. The darker colors denote lower values and brighter colors denote higher values. }\label{fig:distance_mat}
\end{figure}

\begin{table}
    \centering
    \begin{tabular}{p{2cm}p{1.5cm}p{1.5cm}p{1.5cm}p{1.5cm}p{2.5cm}}
    \noalign{\smallskip}\hline\noalign{\smallskip}
                      & Euclidean & Cosine & Correlation & Manhattan & Pearson correlation \\
     \noalign{\smallskip}\svhline\noalign{\smallskip}
    Survey Data       & 9.73      & 0.16   & 0.13        & 65.85     & 0.13  \\
    CF Data & 2.19      & 0.10   & 0.92        & 4.00      & 0.18  \\
    Combined Data   & 6.71      & 0.14   & 0.04        & 45.92     & 0.04  \\ \hline
    \end{tabular}
    \caption{Optimal $\epsilon$ values found for the DBSCAN method using the different features and distance metrics.}
    \label{tab:epsi}
\end{table}

\begin{table}
    \centering
    \begin{tabular}{p{5.5cm}p{5.5cm}}
    \centering
    
    \begin{tabular}{p{2cm}p{2cm}p{1cm}}
    \noalign{\smallskip}\hline\noalign{\smallskip}
                      & \verb|is_neigh| & $k$ \\
     \noalign{\smallskip}\svhline\noalign{\smallskip}
    Survey Data       & 3               & 2 \\
    CF Data & 7               & 2 \\
    Combined Data    & 3               & 2 \\
    \hline
    \end{tabular}  &
    
    \begin{tabular}{p{2cm}p{2cm}p{1cm}}
    \noalign{\smallskip}\hline\noalign{\smallskip}
                      & \verb|n_samples| & $k$ \\
     \noalign{\smallskip}\svhline\noalign{\smallskip}
    Survey Data      & 100             & 25 \\
    CF Data & 150             & 2 \\
    Combined Data     & 100             & 2 \\ \hline
    \end{tabular} \\
    
    \noalign{\smallskip}
    \centering \scriptsize{(a) CkNN Optimal Parameters} & \centering \scriptsize{(b)  CCC Optimal Parameters} \\
    \end{tabular}
    \caption{(a) Optimal parameter values found for the CkNN method using the different features. (b) Optimal parameter values found for the Cut-Cluster-Classify method using the different features.}
    \label{tab:cknn_and_ccc}
\end{table}

\begin{table}
    \centering
    \begin{tabular}{p{2cm}p{1.5cm}p{1.5cm}p{1.5cm}p{3.5cm}}
    \noalign{\smallskip}\hline\noalign{\smallskip}
     Metric & DBSCAN    & Spectral  & CkNN      & Cut-Cluster-Classify (CCC) \\
     \noalign{\smallskip}\svhline\noalign{\smallskip}
   Euclidean & 0.75986 & 0.76534 & 0.76259 & 0.73993\\
      Cosine & 0.75986 & 0.55319 & 0.75540 & 0.75812\\
 Correlation & 0.76190 & 0.62000 & 0.75540 & 0.75986\\
   Manhattan & 0.75986 & 0.62857 & 0.76259 & 0.75540\\
Pearson coeff& 0.7619  & 0.75812 & 0.75540 & 0.76259\\ 
\hline
    \end{tabular}
    \caption{\emph{\textbf{Survey Data}}. F1-scores of using different metrics (rows) to construct the distance matrix. The corresponding distance metric is the input for the different clustering methods (columns). }
    \label{tab:fscores_survey}
\end{table}

\begin{table}[]
    \centering
    \begin{tabular}{p{2cm}p{1.5cm}p{1.5cm}p{1.5cm}p{3.5cm}}
    \noalign{\smallskip}\hline\noalign{\smallskip}
     Metric & DBSCAN    & Spectral  & CkNN      & Cut-Cluster-Classify (CCC) \\
     \noalign{\smallskip}\svhline\noalign{\smallskip}
   Euclidean & 0.75986 & 0.76534 & 0.78481 & 0.75986\\
      Cosine & 0.75986 & 0.76259 & 0.78481 & 0.75655\\
 Correlation & 0.75986 & 0.71937 & 0.74815 & 0.75986\\
   Manhattan & 0.75986 & 0.57143 & 0.78481 & 0.75090\\
Pearson coeff& 0.75986 & 0.54144 & 0.76259 & 0.75986\\ \hline
    \end{tabular}
    \caption{\emph{\textbf{Craniofacial (CF) Data}}. F1-scores of using different metrics (rows) to construct the distance matrix. The corresponding distance metric is the input for the different clustering methods (columns). }
    \label{tab:fscores_craneo}
\end{table}

\begin{table}[]
    \centering
    \begin{tabular}{p{2cm}p{1.5cm}p{1.5cm}p{1.5cm}p{3.5cm}}
    \noalign{\smallskip}\hline\noalign{\smallskip}
     Metric & DBSCAN    & Spectral  & CkNN      & Cut-Cluster-Classify (CCC) \\
     \noalign{\smallskip}\svhline\noalign{\smallskip}
   Euclidean & 0.75986 & 0.76534 & 0.76259 & 0.75986\\
      Cosine & 0.75986 & 0.58537 & 0.7554  & 0.74453\\
 Correlation & 0.75986 & 0.61386 & 0.7554  & 0.75986\\
   Manhattan & 0.76534 & 0.58937 & 0.76259 & 0.75986\\
Pearson coeff& 0.75986 & 0.75812 & 0.75540 & 0.75812\\ \hline
    \end{tabular}
    \caption{\emph{\textbf{Combined Data}}. F1-scores of using different metrics (rows) to construct the distance matrix. The corresponding distance metric is the input for the different clustering methods (columns). }
    \label{tab:fscores_all}
\end{table}
\clearpage
\bibliographystyle{plain}
\bibliography{RevisedFINALProcManuscript}
\end{document}